\documentclass[runningheads]{llncs}

\usepackage{eccv}

\usepackage{eccvabbrv}

\usepackage[table]{xcolor}
\usepackage{graphicx}
\usepackage{booktabs}
\usepackage{adjustbox}
\usepackage{dblfloatfix}
\usepackage{tikz}
\usepackage{makecell}
\usepackage{caption}
\usepackage{multirow}

\definecolor{Appearance_red}{HTML}{F46C6C}
\definecolor{Geometry_yellow}{HTML}{FFB84C}
\definecolor{Gaussian_green}{HTML}{56CB67}

\usepackage[accsupp]{axessibility}  

\usepackage{orcidlink}

\begin{document}

\title{URHead: A Unified UV-Space Representation for Joint Mesh–3DGS Optimization in Head Avatars}

\titlerunning{URHead}

\author{Seonghak Lee\inst{1}\orcidlink{0009-0001-6897-0201} \and
Junhee Cho\inst{1}\orcidlink{0009-0000-0310-8744} \and
Jisoo Park\inst{1}\orcidlink{0009-0004-0885-9538} \and
Min-Gyu Park\inst{2,3}\orcidlink{0000-0003-1752-150X} \and
Jongmin Lee\inst{1}\orcidlink{0000-0001-9410-027X} \and
Ju Hong Yoon\inst{2,3}\orcidlink{0000-0003-2945-8376} \and
Junseok Kwon$^{\dagger}$\inst{1}\orcidlink{0000-0001-9526-7549}
}

\authorrunning{S.~Lee et al.}

\institute{Chung-Ang University \and
Korea Electronics Technology Institute (KETI)\and
polygom\\
  \textsuperscript{$\dagger$} indicates corresponding author.}

\maketitle
\let\thefootnote\relax\footnotetext{Our code is available at \url{https://github.com/Lseonghak/URHead}}

\begin{center}
\vspace{-0.6cm}
{\includegraphics[width=1.00\linewidth]{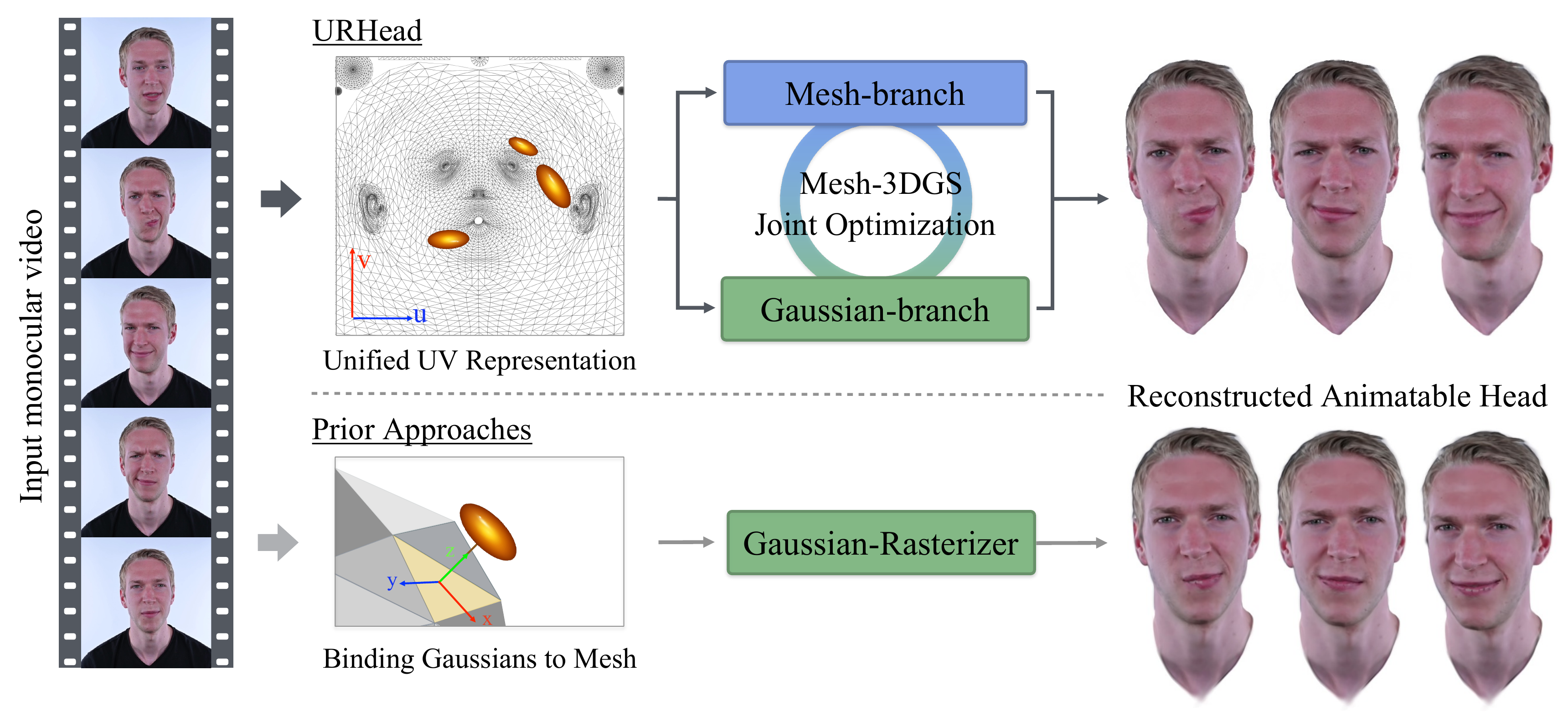}}
    \vspace{-6mm}
    \captionof{figure}{\textbf{Key insight behind URHead}. Unlike prior approaches that bind Gaussians to mesh surfaces in 3D space (bottom), URHead introduces a unified UV-space representation that jointly parameterizes mesh and 3DGS through shared attribute maps (top). Through joint optimization over the shared UV , both representations mutually refine each other, preserving animation controllability while bridging structural coherence and visual fidelity.}
    \label{fig:teaser}
\end{center}

\begin{abstract}
We present URHead, a unified representation for high-fidelity and animatable head avatars that fundamentally redefines mesh-Gaussian integration. While mesh-based methods offer precise geometric control but lack photorealistic detail, and Gaussian-based approaches achieve photorealism but suffer from poor structural consistency, existing hybrid solutions fail to fully leverage their complementary strengths.
Our key contribution is a UV-space unification where both representations share a common UV parameterization. Through joint optimization with adaptive gaussian sampling, our method automatically learns to disentangle and allocate appropriate roles to each component. URHead maintains full parametric controllability while preserving subject-specific details, and outperforms existing state-of-the-art methods in reconstruction quality and animation consistency.
\end{abstract}
\section{Introduction}
\label{sec:intro}
Creating photorealistic digital human avatars has long been a fundamental challenge in computer graphics and vision, with applications spanning virtual reality, film production, telepresence, and gaming~\cite{debevec2012light, guo2019relightables, yang2023towards, cheng2014children, healey2021mixed, kachach2020virtual, li2021vmirror, zackariasson2012video}. Recent advances in neural rendering have dramatically improved the quality of digital humans. 
Nevertheless, current approaches to animatable head avatar reconstruction predominantly rely on either parametric mesh models or 3D Gaussian Splatting (3DGS). Parametric meshes provide explicit geometric structure and controllable deformation, enabling stable animation, whereas 3DGS achieves photorealistic rendering through unstructured Gaussian primitives. However, neither representation alone fully satisfies the demands of both structural coherence and high-frequency detail modeling. 

Parametric facial mesh models~\cite{MICA,DECA,vhap,NPHM,Nonlinear3DFaceMorphableModel} have been widely adopted as a core representation for animatable head avatars, offering explicit topology and controllable deformation through low-dimensional shape and expression parameters. This structured design ensures stable animation and semantic control. However, the same structural priors fundamentally limit geometric expressiveness: the fixed topology cannot faithfully represent high-frequency details such as wrinkles or pores, and the restricted parameter space inevitably discards subject-specific variations outside the learned basis.

In contrast, 3DGS~\cite{3DGS} has emerged as a powerful alternative, representing scenes as anisotropic Gaussian primitives that enable real-time, high-fidelity rendering. However, Gaussians lack explicit topology, making it challenging to maintain structural coherence during animation-driven deformation. To mitigate this issue, recent hybrid approaches~\cite{splattingavatar,gaussianavatar,flashavatar,rgbavatar,gaussianblendshape} incorporate mesh controllability by binding or initializing Gaussians on parametric surfaces. While this introduces coarse geometric guidance, the Gaussians are still optimized independently in 3D space. As a result, over-pruning in sparsely observed regions can remove necessary surface coverage and lead to hollow artifacts, ultimately undermining the structural consistency that mesh integration aims to provide.

Recent studies~\cite{mags, hera, splattingavatar, gaussianavatar, texavatars} have attempted to combine mesh and 3DGS representations, yet most adopt fundamentally decoupled formulations. Methods such as~\cite{splattingavatar, gaussianavatar, mags} constrain Gaussian centers to remain near mesh surfaces to inherit deformation priors, but this spatial binding restricts representational flexibility without enabling true parameter sharing. HERA~\cite{hera} enhances texture fidelity on smooth surfaces, yet maintains separate mesh and Gaussian components optimized independently. 
The core limitation of these hybrid approaches lies in their reliance on 3D-space optimization, where mesh and Gaussians interact only through geometric proximity rather than through a shared parameterization. Consequently, feature exchange between the two remains indirect and incomplete, preventing them from fully leveraging their complementary strengths.

We present URHead, a unified representation for animatable head avatars that \emph{jointly optimizes} mesh and 3D Gaussian primitives under a \emph{common UV parameterization}. Unlike prior hybrid approaches that couple mesh and Gaussians only through spatial constraints in 3D space, URHead reformulates both representations as functions defined over a shared UV domain, enabling true parameter-level integration.
Our key insight is that mesh and Gaussian representations are inherently complementary: meshes provide stable low-frequency geometric structure and controllable deformation, while Gaussians excel at modeling high-frequency residual details and complex appearance variations. Existing approaches either force one representation to approximate all aspects of geometry and appearance~\cite{gaussianavatar, splattingavatar, flashavatar} or combine them through decoupled optimization schemes~\cite{mags,hera}. In contrast, we reformulate both representations as functions defined over a common UV parameterization, enabling true parameter-level unification instead of mere spatial binding.

Within the unified UV space, the mesh provides base geometry and appearance through shared attribute maps, while UV-parameterized Gaussians capture residual details beyond the capacity of fixed-topology meshes. This shared representation enables mesh and Gaussians to be jointly optimized, allowing coordinated refinement of structure and appearance.
Furthermore, UV-space unification naturally supports error-driven adaptive Gaussian allocation. By projecting reconstruction errors onto the surface domain, Gaussians are densified in under-represented regions while redundancy is suppressed elsewhere. Unlike additive blending~\cite{gaussianblendshape,rgbavatar} or surface-constrained spatial binding~\cite{gaussianavatar,splattingavatar,flashavatar}, our approach achieves integration at the representation level.
Through this unified formulation, URHead preserves animation controllability while attaining photorealistic reconstruction quality, bridging structural coherence and visual fidelity. \cref{fig:teaser} illustrates our key insight. We summarize our main contributions as follows:

\begin{itemize}
\item \textbf{Joint UV-space optimization.} To the best of our knowledge, we are the first to jointly optimize mesh and 3D Gaussians within a shared UV-parameterized domain for animatable head avatars.

\item \textbf{Unified UV representation.} We introduce a unified UV-space formulation that integrates mesh geometry and Gaussian attributes within a shared parameterized domain.

\item \textbf{Error-driven Gaussian allocation.} We propose an adaptive Gaussian allocation strategy in UV space for surface-aware densification of under-represented regions.

\item \textbf{Extensive evaluation.} Our approach consistently outperforms mesh-only, Gaussian-only, and prior hybrid methods on standard 3D head avatar benchmarks in reconstruction accuracy and animation stability.
\end{itemize}
\section{Related Work}
\label{sec:related}
\textbf{Mesh representation for head avatars.} 
Mesh representations have long served as the foundation of 3D face modeling due to their explicit topology and efficient rendering. 3D Morphable Models (3DMM) pioneered statistical face modeling by learning PCA bases from 3D scans, parameterizing faces using identity and expression coefficients. Basel Face Model~\cite{BaselFaceModel} expanded the statistical basis with higher quality scans. FaceWarehouse~\cite{FaceWareHouse} introduced multi-linear models for identity and expression. FLAME~\cite{FLAME} incorporated articulation components for jaw and expression control. DECA~\cite{DECA} predicted FLAME parameters along with displacement maps to capture subject-specific geometric details. 
Deferred Neural Rendering~\cite{DefferedNeuralRendering} rasterizes geometry into feature maps that are decoded into view-dependent images. Deep Appearance Models~\cite{Deep_Appearance_Models} assign latent codes to mesh vertices and use view-conditioned MLPs to model spatially-varying BRDF. NeuTex~\cite{NeuTex} aggregates multi-view features into a consistent UV texture space via differentiable sampling, while Neural Head Avatars~\cite{NeuralHeadAvatar} define neural textures on the FLAME topology for avatar rendering.

\noindent \textbf{Hybrid representations for head avatars.}
Recent advances in neural rendering have explored various hybrid representations that combine explicit and implicit modeling.
NeRF~\cite{mildenhall2021nerf} integrated neural fields with traditional graphics primitives. NerFace~\cite{NerFace} combined neural radiance fields with parametric face models\cite{thies2016face2face,blanz2023morphable} for controllable synthesis, while PointAvatar~\cite{Pointavatar} used neural point clouds attached to FLAME mesh for dynamic head modeling. Implicit-explicit hybrids like NPHM~\cite{NPHM} combined the parametric space of FLAME with neural implicit signed distance functions. 
IMAvatar~\cite{IMAvatar} learned implicit fields on top of FLAME geometry to model detailed facial appearance.
Among these hybrid approaches, 3D Gaussian Splatting~\cite{3DGS} has emerged as a particularly powerful representation, modeling scenes as anisotropic Gaussian primitives that enable real-time rendering without neural network queries. The explicit formulation of Gaussians and their success in dynamic modeling~\cite{Dynamic_3DGS, 4DGS} have inspired their integration with parametric models, leading to Mesh–GS hybrid methods.

\noindent \textbf{Mesh-3DGS representation for head avatars.}
Recent studies have explored hybridizing multiple 3D representations to leverage their complementary strengths in expressiveness, efficiency, and controllability.
SplattingAvatar~\cite{splattingavatar} embedded 3D Gaussians on mesh triangles using barycentric coordinates and normal-based displacement, deriving pose-dependent transformations from mesh deformation. GaussianAvatars~\cite{gaussianavatar} introduced rigged 3DGS by binding Gaussians to a FLAME mesh, enabling deformation driven by facial expression and pose parameters~\cite{FLAME}. FlashAvatar~\cite{flashavatar} allocated Gaussians in UV space and applies expression-dependent offsets via an MLP conditioned on FLAME codes, while MonoGaussianAvatar~\cite{monogaussianavatar} learned adaptive Gaussian transformations through deformation fields for expression-aware animation. GaussianBlendShapes~\cite{gaussianblendshape} and RGBAvatar~\cite{rgbavatar} constructed blendshapes directly in Gaussian space, enabling efficient real-time control through linear blending.
While these methods anchor Gaussians to mesh geometry to inherit deformation priors, FATE~\cite{fate} focused on efficient Gaussian utilization via importance-based sampling and post-processing neural baking for improved texture editability.

\noindent \textbf{Positioning of our work.}
In contrast to prior hybrid approaches, URHead achieves parameter-level unification by jointly optimizing mesh and Gaussians within a shared UV-parameterized representation.

\begin{figure*}[!t]
\centering
\includegraphics[width=1.0\linewidth]{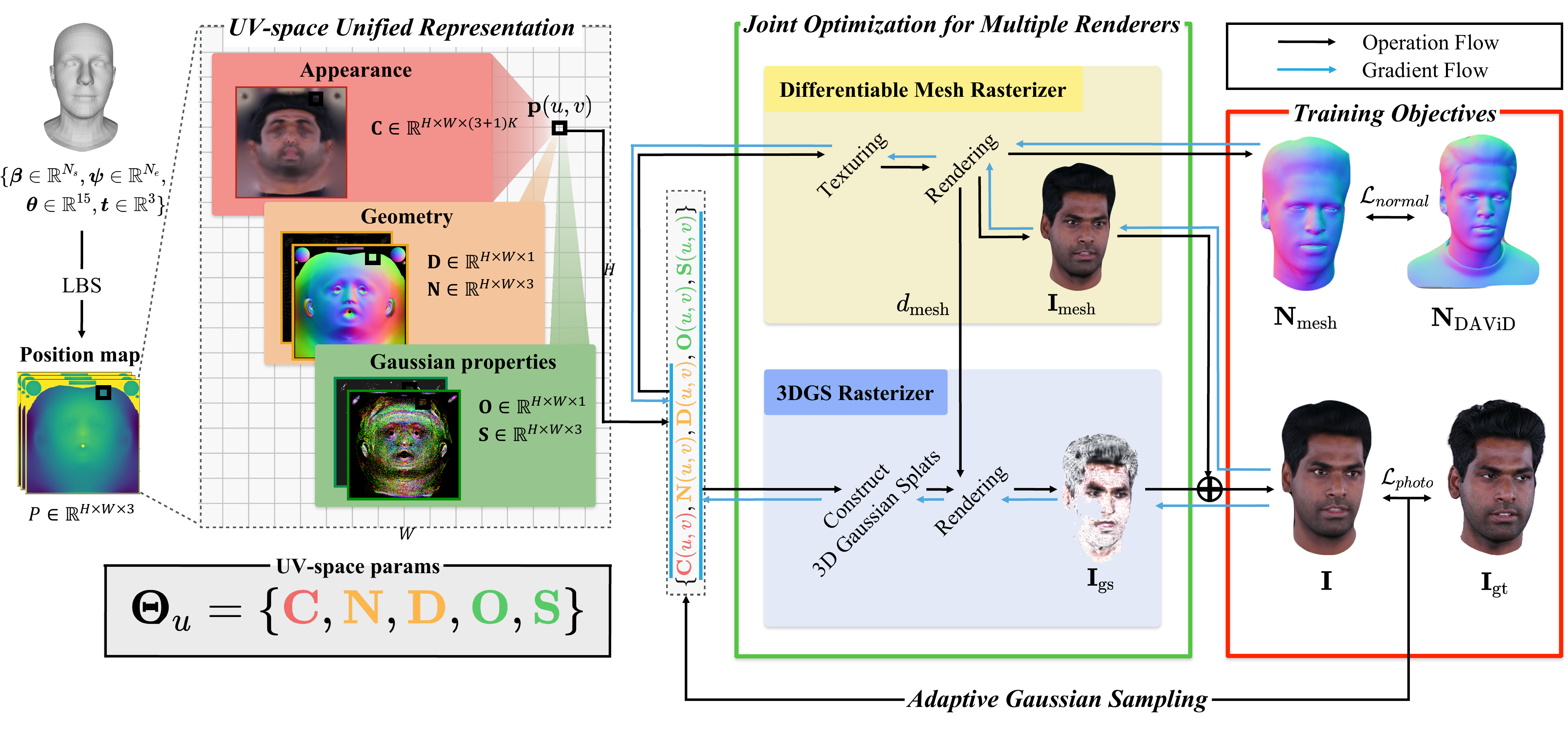}
\vspace{-7mm}
\caption{\textbf{Overall pipeline of URHead.} Given monocular video frames and FLAME parameters ($\boldsymbol{\beta}$, $\boldsymbol{\psi}$, $\boldsymbol{\theta}$, $\boldsymbol{t}$), we construct a position map $\mathbf{P}$ through linear blend skinning (LBS)~\cite{FLAME}. Our unified UV-space representation $\boldsymbol{\Theta}_{u} = \{{\color{Appearance_red}\mathbf{C}}, {\color{Geometry_yellow}\mathbf{N}}, {\color{Geometry_yellow}\mathbf{D}}, {\color{Gaussian_green}\mathbf{O}}, {\color{Gaussian_green}\mathbf{S}}\}$ consists of {\color{Appearance_red}appearance}, {\color{Geometry_yellow}geometry}, and {\color{Gaussian_green}gaussian-specific} parameters. 
The joint optimization framework employs two renderers: a differentiable mesh rasterizer takes $\{ \mathbf{C}, \mathbf{N}, \mathbf{D}\}$, and produces $\mathbf{I}_{\text{mesh}}$, $\mathbf{N}_{\text{mesh}}$, and depth $d_{\text{mesh}}$. Meanwhile, the 3DGS rasterizer takes $\{\mathbf{C}, \mathbf{N}, \mathbf{D}, \mathbf{O}, \mathbf{S}\}$, and generates $\mathbf{I}_{\text{gs}}$ using occlusion-aware rendering. Both renderers share the UV-space parameters and are jointly optimized through gradient flows (blue arrows), enabling mutual refinement. The final image $\mathbf{I}$ is composited via alpha blending. Training objectives include photometric loss $\mathcal{L}_{\text{photo}}$ and normal consistency loss $\mathcal{L}_{\text{normal}}$ supervised by pseudo ground-truth $\mathbf{N}_{\text{DAViD}}$. The Adaptive Gaussian Sampling module allocates new points based on reconstruction error.}
    
\label{fig:pipeline}
\vspace{-5mm}
\end{figure*}

\section{Proposed Method}
\vspace{-2mm}
The proposed URHead consists of four main components. 
\cref{sec:unified_uv} introduces the unified UV-space facial representation that leverages the complementary properties of mesh and 3DGS. 
\cref{sec:optimize_multi_render} describes the joint optimization framework that aligns multiple renderers under this unified formulation. 
\cref{3_proposed:error-based sampling} presents an adaptive Gaussian sampling strategy for efficient UV-space allocation. 
\cref{sec:training_objectives} details the training objectives and overall loss function. 
\cref{fig:pipeline} illustrates the complete URHead pipeline.

\vspace{-4mm}
\subsection{Unified UV-Parameterized Representation} \label{sec:unified_uv}
\vspace{-2mm}
Our unified UV-space representation allows structured geometry and fine-grained appearance to be jointly encoded and optimized under a common parameterization. By reformulating both mesh and Gaussian attributes in UV space, we overcome their complementary limitations: meshes are restricted by fixed topology in capturing high-frequency detail, whereas 3DGS lack structural constraints for stable deformation.

Our framework builds upon a parametric facial model~\cite{FLAME, NPHM, Nonlinear3DFaceMorphableModel} that provides a consistent UV parameterization with controllable facial expressions and poses. In our implementation, we adopt FLAME~\cite{FLAME}, though other parametric facial models could be used.
FLAME defines facial parameters as $\boldsymbol{\Theta}_{f}=\{\boldsymbol{\beta}, \boldsymbol{\psi}, \boldsymbol{\theta}, \boldsymbol{t}\}$, where $\boldsymbol{\beta} \in \mathbb{R}^{N_s}$ encodes identity shape, $\boldsymbol{\psi} \in \mathbb{R}^{N_e}$ models facial expressions, $\boldsymbol{\theta} \in \mathbb{R}^{15}$ represents pose parameters (global rotation, neck, jaw, and eyeballs), and $\boldsymbol{t} \in \mathbb{R}^3$ denotes translation. Given $\boldsymbol{\Theta}_{f}$, the model generates a head mesh with vertices $\mathcal{V}$ and faces $\mathcal{F}$, together with a UV parameterization that maps each vertex to texture coordinates.
This UV mapping establishes a shared parameterized domain, enabling appearance, geometry, and Gaussian attributes to be represented consistently in dense UV maps.

Leveraging the shared UV parameterization, we formulate a unified representation in a common parameterized domain:
\begin{equation}
\boldsymbol{\Theta}_{u} = \{\mathbf{C}, \mathbf{N}, \mathbf{D}, \mathbf{O}, \mathbf{S}\},
\end{equation}
where all facial attributes are expressed in a shared UV-parameterized domain as dense maps. 
The representation is composed of two groups: shared attribute maps $(\mathbf{C}, \mathbf{N}, \mathbf{D})$, jointly utilized by both mesh and Gaussian renderers, and Gaussian-specific maps $(\mathbf{O}, \mathbf{S})$, which encode properties unique to Gaussian primitives.
The color map $\mathbf{C} \in \mathbb{R}^{H \times W \times (3+1)K}$ stores spherical harmonics (SH) coefficients for view-dependent appearance, where $K=(d+1)^2$ with SH degree $d$. The normal map $\mathbf{N} \in \mathbb{R}^{H \times W \times 3}$ encodes per-pixel surface orientation in UV space, supporting both mesh shading and Gaussian rotation. The displacement map $\mathbf{D} \in \mathbb{R}^{H \times W}$ introduces learnable geometric details by displacing vertices along surface normals.
These shared maps $(\mathbf{C}, \mathbf{N}, \mathbf{D})$ are jointly utilized by both the mesh and Gaussian Splatting (GS) renderers.
For Gaussian rendering, we additionally define an opacity map $\mathbf{O} \in \mathbb{R}^{H \times W}$ and a scale map $\mathbf{S} \in \mathbb{R}^{H \times W \times 3}$. The opacity map controls the transparency of each Gaussian primitive, while the scale map determines its anisotropic extent along three axes. Gaussian rotation is derived from the shared normal map $\mathbf{N}$ via quaternion conversion, ensuring alignment with local surface geometry.

\subsection{Joint UV-Space Optimization of Mesh and 3DGS}\label{sec:optimize_multi_render}
We jointly optimize the mesh and Gaussians to combine their complementary strengths. 
Optimizing them independently often leads to misaligned geometry and residual details, 
so a shared UV-space optimization maintains consistent updates across both representations.

The mesh renderer recovers coarse geometry and smooth appearance from the FLAME template~\cite{FLAME}, while the Gaussian renderer supplements high-frequency residual details that are difficult to capture with a template mesh alone. 
Given parametric inputs $\boldsymbol{\Theta}_{f}$, FLAME produces base vertices $\mathcal{V} = \{\mathbf{v}_i\}_{i=1}^{N_v}$ for each frame. We rasterize the deformed mesh into UV space to construct a position map $\mathbf{P} \in \mathbb{R}^{H \times W \times 3}$, forming a dense 3D position field. For any UV coordinate $(u,v)$, the corresponding 3D position is defined as
\begin{equation}
\mathbf{x}(u,v) = \mathbf{P}(u, v) + \mathbf{D}(u, v) \cdot \mathbf{N}(u, v),
\label{rq:position}
\end{equation}
where $\mathbf{P}(u,v)$ denotes the base mesh position and $\mathbf{D}(u,v)$ introduces learnable displacement along the normal $\mathbf{N}(u,v)$. As $\mathbf{P}$ is updated per frame from the deformed mesh, consistent 3D–UV correspondence is preserved across renderers.

\noindent \textbf{Mesh Rendering.}
We obtain displaced vertices $\hat{\mathcal{V}}$ by sampling the displacement map $\mathbf{D}$ at each vertex UV coordinate $(u_i, v_i)$:
\[
\hat{\mathbf{v}}_i = \mathbf{v}_i + \mathbf{D}(u_i, v_i)\cdot \mathbf{N}(u_i, v_i),
\]
where $\mathbf{v}_i$ denotes the FLAME vertex and $\mathbf{N}$ is the shared normal map. 
The displaced mesh is rendered with nvdiffrast~\cite{Laine2020diffrast} via
\begin{equation}  \label{eq:mesh}
\mathbf{I}_{\text{mesh}}, \mathbf{N}_{\text{mesh}}, d_{\text{mesh}} = \mathcal{R}_{\text{mesh}}(\mathbf{C}, \mathbf{N}, \hat{\mathcal{V}}, \mathcal{F}),
\end{equation}
producing the rendered mesh image $\mathbf{I}_{\text{mesh}}$, rendered normal map $\mathbf{N}_{\text{mesh}}$, and mesh depth map $d_{\text{mesh}}$. 
The shared UV maps $\mathbf{C}$ and $\mathbf{N}$ are sampled using per-pixel UV coordinates from rasterization.
To better capture geometric displacement, we refine $\mathbf{N}$ using supervision from the monocular normal estimator DAViD~\cite{david}, which provides pseudo ground-truth normals. 
The resulting depth map $d_{\text{mesh}}$ is later used for occlusion-aware Gaussian rendering.

\noindent\textbf{Gaussian Rendering.}
Unlike standard 3D Gaussian Splatting~\cite{3DGS}, where Gaussian centers are directly defined in 3D space, we parameterize Gaussians in UV space as $\mathcal{G} \in \mathbb{R}^{N_{gs} \times 2}$. 
For each UV coordinate $(u_j, v_j) \in \mathcal{G}$, the corresponding 3D position $\mathbf{p}$ is obtained by lifting the 2D location through the position map $\mathbf{P}$. 
Gaussian rotation is derived from the shared normal map $\mathbf{N}$ via quaternion transformation, while opacity, scale, and color are sampled from the UV maps $\mathbf{O}$, $\mathbf{S}$, and $\mathbf{C}$, respectively.

To account for visibility interactions with the mesh, we modify the 3DGS rasterizer~\cite{3DGS} for occlusion-aware rendering:
\begin{equation} \label{eq:gs}
\mathbf{I}_{\text{gs}}, T = \mathcal{R}_{\text{gs}}(\mathbf{C}, \mathbf{N}, \mathbf{O},
\mathbf{S}, d_{\text{mesh}},\mathcal{G}),
\end{equation}
where $\mathbf{I}_{\text{gs}}$ is the rendered Gaussian image, $\mathcal{G}$ represents the lifted 3D Gaussians, and $d_{\text{mesh}}$ is the mesh depth map. Gaussians behind the mesh surface ($z_j > d_{\text{mesh}}(x,y) + \epsilon$, with $\epsilon=0.01$) are culled, and visible Gaussians are composited via alpha blending. 
The rasterizer outputs the transmittance $T$, representing the accumulated visibility of the Gaussian layer and regulating the contribution of the mesh layer in the final composition.

\noindent\textbf{Joint Optimization of Mesh and 3DGS.}
Because the mesh and Gaussian renderers share the UV-space attribute maps $(\mathbf{C}, \mathbf{N}, \mathbf{D})$, gradients from both rendering paths jointly update these parameters during backpropagation. Taking the normal map $\mathbf{N}$ as an example, the total gradient decomposes as
\begin{equation}
\frac{\partial \mathcal{L}_{\text{total}}}{\partial \mathbf{N}} = \underbrace{\frac{\partial \mathcal{L}_{\text{photo}}}{\partial \mathbf{I}_{\text{mesh}}} \frac{\partial \mathbf{I}_{\text{mesh}}}{\partial \mathbf{N}}}_{\text{Mesh Flow}} + \underbrace{\frac{\partial \mathcal{L}_{\text{photo}}}{\partial \mathbf{I}_{\text{gs}}} \frac{\partial \mathbf{I}_{\text{gs}}}{\partial \mathbf{N}}}_{\text{Gaussian Flow}},
\label{eq:total flow}
\end{equation}
revealing explicit cross-renderer gradient coupling. Since mesh vertices are displaced along $\mathbf{N}$ in \cref{eq:total flow}, the Gaussian flow directly refines mesh geometry through the shared normal map. The same gradient structure applies to $\mathbf{C}$ and $\mathbf{D}$, enabling coordinated refinement between mesh and 3DGS.
The losses $\mathcal{L}_{\text{photo}}$ and $\mathcal{L}_{\text{total}}$ are defined in \cref{eq:L_photo} and \cref{eq:L_total}, respectively.

\noindent \textbf{Image Composition.}
The final image is obtained by compositing the Gaussian and mesh layers via alpha blending~\cite{3DGS}:
\begin{equation} \label{eq:composites}
{{\mathbf{I}}} = \mathbf{I}_{\text{gs}} + T \cdot \mathbf{I}_{\text{mesh}},
\end{equation}
where $\mathbf{I}_{\text{gs}}$ in \cref{eq:gs} contributes high-frequency details in the front, and $\mathbf{I}_{\text{mesh}}$ in \cref{eq:mesh} provides the base geometry weighted by the transmittance $T$. This formulation allows Gaussians to refine mesh rendering by adding residual details while preserving the underlying mesh structure.

\begin{figure} [t]
    \centering
    \begin{tikzpicture}
    \node[anchor=south west,inner sep=0] (img) at (0,0){\includegraphics[width=1.0\linewidth]{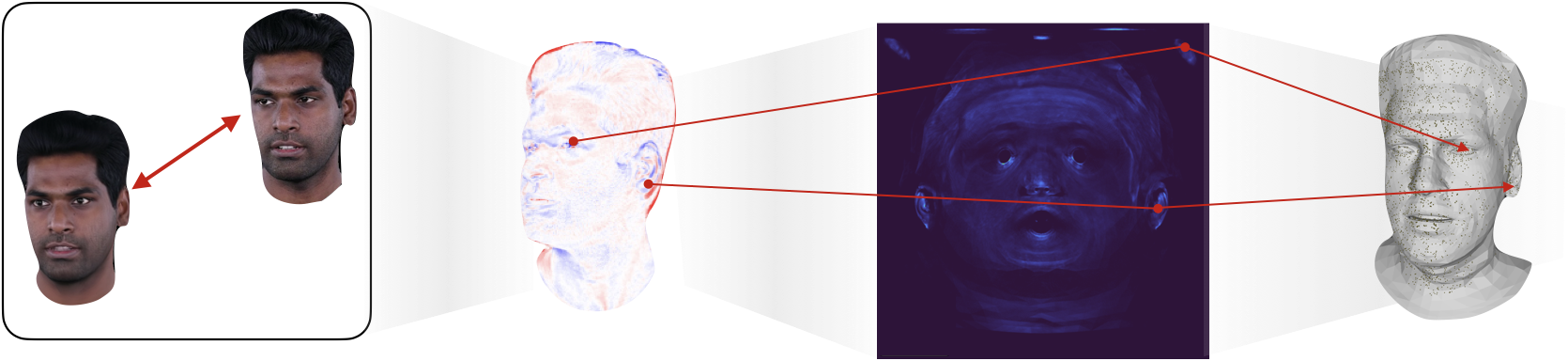}};
    \node[anchor=south] at (0.05\textwidth, 0.00\textwidth) {\strut \scriptsize $\mathbf{I}$};
    \node[anchor=south] at (0.125\textwidth, 0.11\textwidth) {\strut \scriptsize ${L}_\text{1}$};
    \node[anchor=south] at (0.20\textwidth, 0.06\textwidth) {\strut \scriptsize $\mathbf{I}_\text{gt}$};
    \node[anchor=south] at (0.39\textwidth, -0.015\textwidth) {\strut \scriptsize $\mathbf{E}_\text{img}$};
    \node[anchor=south] at (0.125\textwidth, -0.04\textwidth) {\strut \scriptsize \textbf{Estimating error}};
        \node[anchor=south, rotate=90] at (0.52\textwidth, 0.115\textwidth) {\strut \tiny \textbf{UV Sampling}};
    \node[anchor=south] at (0.65\textwidth, -0.04\textwidth) {\strut \scriptsize \textbf{UV Error map}};
    \node[anchor=south] at (0.93\textwidth, -0.04\textwidth) {\strut \scriptsize \textbf{3D position}};
    \vspace{-3mm}
    \end{tikzpicture}
        \caption{\textbf{Adaptive Gaussian allocation through error-driven sampling.} We project photometric reconstruction errors from image space to UV space, producing an error map (second column) where brighter colors indicate larger errors. Based on this distribution, we sample new Gaussians, shown as 3D point clouds (last column).}
    \label{fig:sampling}
        \vspace{-3mm}   
\end{figure}
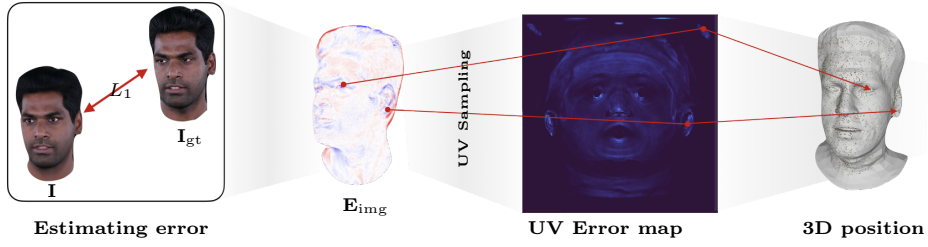

\subsection{Error-Driven Adaptive Gaussian Sampling in UV Space}
\label{3_proposed:error-based sampling}
Our shared UV-parameterized domain provides a surface-aligned space consistent across all renderers. This unified domain naturally enables adaptive Gaussian sampling, as reconstruction errors can be accumulated directly on the UV surface to guide where additional Gaussians are required. \cref{fig:sampling} illustrates the resulting sampling distribution and adaptive allocation.
Unlike standard 3D Gaussian Splatting~\cite{3DGS}, which performs gradient-based densification in unstructured 3D space, we sample new Gaussians directly in UV space. 
Operating in the shared UV domain allows photometric reconstruction errors to be projected back to their corresponding surface locations. 
We compute the per-pixel photometric error as $\mathbf{E}_{\text{img}} = \| \mathbf{I} - \mathbf{I}_{\text{gt}} \|_1 \in \mathbb{R}^{H \times W}$ and simultaneously render aligned UV coordinates to map these errors onto the UV surface. 
Regions with high reconstruction error typically indicate missing high-frequency details (\eg wrinkles, pores, subtle facial structures) and thus require additional Gaussian primitives.

Given an input video sequence, we accumulate UV-space reconstruction errors over time and normalize them as
\begin{equation}
w(u,v) = \frac{1}{N}{\sum_{t \in T} V(u, v, t) \mathbf{E}_{\text{uv}}(u, v, t)},
\end{equation}
where $V(u, v, t) \in \{0,1\}$ is a visibility indicator at frame $t$\footnote{For notational simplicity, we omit $t$ in other sections. It is also omitted in $p(u,v)$, as the normalized error is accumulated over the entire sequence.}, equal to 1 if visible and 0 otherwise.
$\mathbf{E}_{\text{uv}}(u, v, t)$ denotes the UV-projected photometric error derived from the image-space error $\mathbf{E}_{\text{img}}$, and $N$ is the number of visible observations.
Higher values of $w(u,v)$ indicate low reconstruction confidence and thus higher sampling probability. Accordingly, we allocate new Gaussian primitives in proportion to the accumulated error to compensate for the template mesh's limited capacity to represent fine details. 
To maintain stable optimization, we cap the number of newly added Gaussians per iteration and prune primitives with low opacity or excessive screen-space extent following~\cite{3DGS}. Newly initialized Gaussians sample their attributes from $\boldsymbol{\Theta}_u$ and are appended to $\mathcal{G}$.

\subsection{Training Objectives}\label{sec:training_objectives}

We optimize URHead using a combination of structural, photometric, normal, and geometric regularization losses. The total objective is defined as
\begin{equation} \label{eq:L_total}
\mathcal{L}_{\text{total}} = \lambda_{\text{ssim}}\mathcal{L}_{\text{ssim}} + (1-\lambda_{\text{ssim}})\mathcal{L}_{\text{photo}} + \lambda_\text{n}\mathcal{L}_{\text{normal}} +
\lambda_\text{L}\mathcal{L}_{\text{lap}}, 
\end{equation}
where $\mathcal{L}_{\text{photo}}$ denotes photometric reconstruction, $\mathcal{L}_{\text{normal}}$ enforces normal consistency, and $\mathcal{L}_{\text{lap}}$ represents Laplacian geometric regularization. The weights $\lambda_\text{n}$ and $\lambda_\text{L}$ balance normal supervision and geometric smoothness, respectively.

The photometric loss consists of $\ell_1$ reconstruction and SSIM\cite{ssim} terms applied to both the composited image and the mesh-only rendering:
\begin{equation}
\begin{split}
\mathcal{L}_{\text{photo}} = &\ 0.8 \cdot \|{\mathbf{I}} - \mathbf{I}_{\text{gt}}\|_1 
+ 0.2 \cdot \|\mathbf{I}_{\text{mesh}} - \mathbf{I}_{\text{gt}}\|_1,
\end{split}\label{eq:L_photo}
\end{equation}
\begin{equation}
\mathcal{L}_{\text{ssim}} = \text{SSIM}({\mathbf{I}}, \mathbf{I}_{\text{gt}}), 
\end{equation}
where $\mathbf{I}$ in~\cref{eq:composites} denotes the final composited image and $\mathbf{I}_{\text{mesh}}$ in \cref{eq:mesh} is the mesh-rendered image.
The mesh-only term encourages the template mesh to capture coarse appearance and global texture, while the composited image terms guide high-frequency detail reconstruction and overall visual fidelity.
The normal supervision loss is defined as
\begin{equation}
\mathcal{L}_{\text{normal}} = \|\mathbf{N}_{\text{mesh}} - \mathbf{N}_{\text{DAViD}}\|_1,
\end{equation}
where $\mathbf{N}_{\text{DAViD}}$ is a pseudo ground-truth normal map estimated by the off-the-shelf method DAViD~\cite{david}. This term encourages accurate geometric displacement and surface orientation.
To suppress noisy displacements, we apply Laplacian mesh regularization:
\begin{equation}
\mathcal{L}_{\text{lap}} = \frac{1}{N} \sum_{i=1}^{N} \|\boldsymbol{\delta}_i\|^2,
\end{equation}
where $\boldsymbol{\delta}_i = \mathbf{v}_i - \frac{1}{|\mathcal{A}_i|} \sum_{j \in \mathcal{A}_i} \mathbf{v}_j$ denotes the Laplacian coordinate of vertex $\mathbf{v}_i$, measuring its deviation from the mean of its adjacent vertices $\mathcal{A}_i$.

\section{Experimental Results}
We conducted extensive experiments on public benchmark datasets, evaluating both quantitative and qualitative results from multiple perspectives. In addition, we analyzed the characteristics of our method through ablation studies.
\textit{More results are provided in the supplementary materials.}

\vspace{-6mm}
\subsection{Experimental Setups}\label{sec:imple_details}
\vspace{-2mm}

\noindent\textbf{Training Details.}
We used a UV resolution of $2048 \times 2048$ for all UV maps. The FLAME model was configured with $N_s = 300$ identity shape coefficients and $N_e = 100$ expression coefficients. For spherical harmonics, we used degree 3 ($K = 16$ coefficients per color channel). The color and normal maps were initialized from VHAP~\cite{vhap} tracking. Gaussians with opacity below $0.01$ were pruned during densification, and the occlusion threshold for depth-aware rendering was set to $\epsilon = 0.01$. All models were trained for 50,000 iterations on a single NVIDIA RTX 3090 GPU. The training and rendering times are 120 mins and 62 FPS, respectively.

We used the Adam optimizer\cite{adam} and the learning rates are $\{5{\times}10^{-4}, 10^{-3}$, $5{\times}10^{-3}, 5{\times}10^{-2}, 10^{-5}\}$ for UV maps $(\mathbf{C},\mathbf{N},\mathbf{S},\mathbf{O},\mathbf{D})$, and $\{10^{-3}, 10^{-5}, 10^{-6}\}$ for FLAME $(\boldsymbol{\psi}, \boldsymbol{\theta}, \boldsymbol{t})$ to ensure training stability. An exponential learning rate decay was applied to the UV parameters after the initial training iterations. Following~\cite{gem}, we evaluated all methods using a head-region mask that matches the FLAME topology for fair comparison.


\noindent
\textbf{Datasets.}
We evaluated our model on three monocular head reconstruction datasets: INSTA~\cite{INSTA}, PointAvatar~\cite{Pointavatar}, and NerFace~\cite{NerFace}. All video frames were preprocessed using Robust Video Matting (RVM)~\cite{rvm} for background removal. Each frame was resized to $512\times 512$ resolution. For training and evaluation, we split the frames into two subsets: all frames except the last 350 were used for training, while the final 350 frames were reserved for testing. 
Our preprocessing followed that of GaussianAvatars~\cite{gaussianavatar}, which employs VHAP\cite{vhap} for FLAME parameter tracking, while for other baseline methods, we adopted their respective preprocessing procedures to ensure optimal performance for each approach. 

\noindent\textbf{Baselines.}
We compared our method with six monocular Gaussian-based head avatar methods: SplattingAvatar~\cite{splattingavatar}, GaussianAvatars~\cite{gaussianavatar}, FlashAvatar~\cite{flashavatar}, GaussianBlendShape~\cite{gaussianblendshape}, FateAvatar~\cite{fate}, and RGBAvatar~\cite{rgbavatar}. These methods represent diverse strategies for integrating Gaussians with parametric meshes, including triangle binding~\cite{splattingavatar, gaussianavatar}, UV-space allocation~\cite{flashavatar}, and expression blendshapes in Gaussian space~\cite{gaussianblendshape, rgbavatar}.

\definecolor{best}{RGB}{255, 142, 142}
\definecolor{second}{RGB}{255, 204, 153}
\begin{table*}[t]
\centering
\caption{\textbf{Quantitative comparisons}. We compare our approach with six state-of-the-art methods. The \textcolor{best}{best} and \textcolor{second}{second} results are highlighted. Abbreviations indicate SplattingAvatar (SA)~\cite{splattingavatar}, GaussianAvatars (GA)~\cite{gaussianavatar}, FlashAvatar (FA)~\cite{flashavatar}, GaussianBlendShape (GBS)~\cite{gaussianblendshape},  FateAvatar (FATEA)~\cite{fate}, and RGBAvatar (RGBA)~\cite{rgbavatar}.
\vspace{-5mm}
\label{table:INSTA_and_GaussianBlendshape}}
\scriptsize
\setlength{\tabcolsep}{5pt}
\renewcommand{\arraystretch}{1.3}
\begin{adjustbox}{max width=\textwidth}
\begin{tabular}{c|c|ccccc|c|cc|c|ccc|c}
\hline
\multicolumn{2}{c|}{\multirow{2}{*}{Datasets}} &
\multicolumn{6}{c|}{INSTA Dataset~\cite{INSTA}} &
\multicolumn{3}{c|}{PointAvatar Dataset~\cite{Pointavatar}} &
\multicolumn{4}{c}{NerFace Dataset~\cite{NerFace}} \\
\cline{3-15}
\multicolumn{2}{c|}{} & bala & biden & justin & malte\_l &  wojtek\_1 & avg. & yufeng& marcel& avg.& nf\_01& nf\_02& nf\_03& avg. \\
\hline
\multirow{7}{*}{ \makecell{PSNR \\ ($\uparrow$) }} 
& SA~\cite{splattingavatar}     & 32.6698 & 29.8051 & 34.9910 & 27.4599 & 32.9244 & 31.5700 &  26.8354& 24.5210& 25.6782& 28.4617 & 31.0630 & 29.8073 & 29.7773 \\
& GA~\cite{gaussianavatar}      & 33.9773 & 31.8956 & 35.9398 & 32.0374& 33.0825 & 33.3865& 25.3688& 26.2412 & 25.8050& 29.5514 & 33.6670 & 29.9414 & 31.0533 \\
& FA~\cite{flashavatar}         & 32.7987 & 31.1547 & 35.5084 & 30.1380 & 32.4582 & 32.4116 &  29.3201& 26.5384& 27.9293& 20.4025 & 32.7650 & 20.5508 & 24.5728 \\
& GBS~\cite{gaussianblendshape} & 34.7045& \cellcolor{second}31.9907& \cellcolor{best}37.0975& 31.6660& 34.7700 & 34.0457 & \cellcolor{second}31.0313& 24.6815 & 27.8564& 30.4162 & 32.8143 & 29.7572 & 30.9959 \\
& FATEA~\cite{fate} & 33.6228& 30.6082& 34.7205& 30.4780& 34.2138 & 32.7287& 29.9326& 24.7649  & 27.3488& 29.5284 & 32.9680 & 29.0570 & 30.5178 \\
& RGBA~\cite{rgbavatar}         & \cellcolor{second}35.1581& 31.9064& \cellcolor{second}36.7658 & \cellcolor{second}31.6830&  \cellcolor{second}35.2092 & \cellcolor{second}34.1445 & \cellcolor{best}31.4727&\cellcolor{second}25.4436 & \cellcolor{second}28.4582& \cellcolor{best}30.6080 & \cellcolor{best}34.8521 & \cellcolor{best}30.4110 & \cellcolor{best}31.9570 \\
& Ours & \cellcolor{best}35.3185& \cellcolor{best}34.3604& 36.5552& \cellcolor{best}34.0563 &\cellcolor{best}35.6023 & \cellcolor{best}35.1785  & 26.6888&   \cellcolor{best}29.7056& \cellcolor{best}28.1972& \cellcolor{second}30.4711 & \cellcolor{second}34.4519 & \cellcolor{second}29.9626 & \cellcolor{second}31.6285 \\
\hline
\multirow{7}{*}{\makecell{SSIM \\ ($\uparrow$)}} 
& SA~\cite{splattingavatar}     & 0.9325& 0.9522& 0.9623& 0.9294&0.9591& 0.9471 & 0.9052& 0.9253& 0.9153& 0.9306 & 0.9565 & 0.9296 & 0.9389 \\
& GA~\cite{gaussianavatar}      & 0.9445& 0.9681& 0.9704& 0.9632& 0.9632& 0.9619  & 0.9054& 0.9373& 0.9214& 0.9407 & 0.9606 & 0.9356 & 0.9456 \\
& FA~\cite{flashavatar}         & 0.9250& 0.9587& 0.9654& 0.9451& 0.9570 &  0.9502& 0.9291& 0.9352& 0.9322& 0.9088 & 0.9579 & 0.8954 & 0.9207 \\
& GBS~\cite{gaussianblendshape} & \cellcolor{second}0.9563& 0.9693& \cellcolor{best}0.9753& \cellcolor{second}0.9633& \cellcolor{best}0.9728 & \cellcolor{second}0.9674 & \cellcolor{second}0.9421& 0.9359& \cellcolor{second}0.9390& \cellcolor{best}0.9520 & 0.9573 & \cellcolor{second}0.9376 & 0.9490 \\
& FATEA~\cite{fate}             & 0.9365& 0.9617& 0.9628& 0.9505& 0.9663& 0.9556 & 0.9303& 0.9346 & 0.9325& 0.9403 & 0.9595 & 0.9212 & 0.9403 \\
& RGBA~\cite{rgbavatar}         & 0.9541& \cellcolor{best}0.9718& 0.9722& 0.9523& 0.9663 & 0.9633& \cellcolor{best}0.9460& \cellcolor{second}0.9379& \cellcolor{best}0.9420& 0.9483 & \cellcolor{second}0.9678 & \cellcolor{best}0.9482 & \cellcolor{best}0.9548 \\
& Ours & \cellcolor{best}0.9585& \cellcolor{second}0.9702& \cellcolor{second}0.9725& \cellcolor{best}0.9649& \cellcolor{second}0.9716 & \cellcolor{best}0.9675 & 0.9135& \cellcolor{best}0.9520& 0.9328& \cellcolor{second}0.9490 & \cellcolor{best}0.9726 & 0.9313 & \cellcolor{second}0.9510 \\
\hline
\multirow{7}{*}{\makecell{LPIPS \\ ($\downarrow$)}} 
& SA~\cite{splattingavatar}     & 0.0432& 0.0415& 0.0333& 0.0424& 0.0261& 0.0373& 0.0578& 0.0844& 0.0711& 0.0583 & 0.0211 & 0.0627 & 0.0474 \\
& GA~\cite{gaussianavatar}      & 0.0537& 0.0281& 0.0239& 0.0333& 0.0241& 0.0326 & 0.0848& 0.0723& 0.0785& 0.0533 & 0.0273 & 0.0562 & 0.0456 \\
& FA~\cite{flashavatar}         & 0.1044& 0.0593& 0.0411& 0.0681& 0.0280& 0.0602 & 0.0471& 0.0820& 0.0645& 0.1120 & 0.0223 & 0.1393 & 0.0912 \\
& GBS~\cite{gaussianblendshape} & 0.0380& 0.0394& 0.0247& 0.0398& 0.0211& 0.0326& 0.0497& 0.0756& 0.0626& 0.0537 & 0.0299 & 0.0592 & 0.0476 \\
& FATEA~\cite{fate}             & 0.0431& 0.0303& 0.0239& 0.0347& 0.0201&  0.0304& \cellcolor{best}0.0389& 0.0613& \cellcolor{second}0.0501& 0.0466 & 0.0225 & 0.0549 & 0.0413 \\
& RGBA~\cite{rgbavatar}         & \cellcolor{second}0.0324& \cellcolor{second}0.0255& \cellcolor{second}0.0193& \cellcolor{second}0.0303& \cellcolor{second}0.0178 & \cellcolor{second}0.0251  & \cellcolor{second}0.0410& \cellcolor{second}0.0664& 0.0537& \cellcolor{second}0.0457 & \cellcolor{second}0.0183 & \cellcolor{second}0.0469 & \cellcolor{second}0.0370 \\
& Ours                          & \cellcolor{best}0.0226& \cellcolor{best}0.0123& \cellcolor{best}0.0150& \cellcolor{best}0.0130& \cellcolor{best}0.0117 & \cellcolor{best}0.0149 & 0.0523& \cellcolor{best}0.0392& \cellcolor{best}0.0458& \cellcolor{best}0.0323 & \cellcolor{best}0.0168 & \cellcolor{best}0.0444 & \cellcolor{best}0.0312 \\
\hline
\end{tabular}

\end{adjustbox}
\vspace{-7mm}
\end{table*}

\vspace{-5mm}
\subsection{Quantitative Comparisons}\label{sec:score_results}
\vspace{-2mm}

\cref{table:INSTA_and_GaussianBlendshape} shows the quantitative results on the INSTA~\cite{INSTA}, PointAvatar~\cite{Pointavatar}, and NerFace~\cite{NerFace} datasets. Overall, our method consistently outperforms baseline methods across diverse subjects and datasets.
On the INSTA dataset~\cite{INSTA} with 5 subjects (one subject overlaps with PointAvatar), URHead achieves an average PSNR of 35.1785 (+1.03 over RGBAvatar~\cite{rgbavatar}), SSIM of 0.9675, and LPIPS of 0.0149. Notably, our model achieves the best LPIPS across all subjects compared to the previous best RGBAvatar~\cite{rgbavatar} (0.0251), indicating that our UV-space unification particularly benefits perceptual quality. Our method also improves performance on the PointAvatar and NerFace datasets~\cite{Pointavatar, NerFace}.
The consistent improvements across pixel-level (PSNR), structural (SSIM), and perceptual (LPIPS) metrics confirm that our unified UV-space representation effectively leverages the geometric priors of FLAME~\cite{FLAME} while integrating the complementary strengths of both mesh and Gaussian representations. We also observe lower performance on the \textit{yufeng} subject of PointAvatar~\cite{Pointavatar}, whose long hair extends beyond the FLAME topology; we discuss this limitation further in the supplementary material.

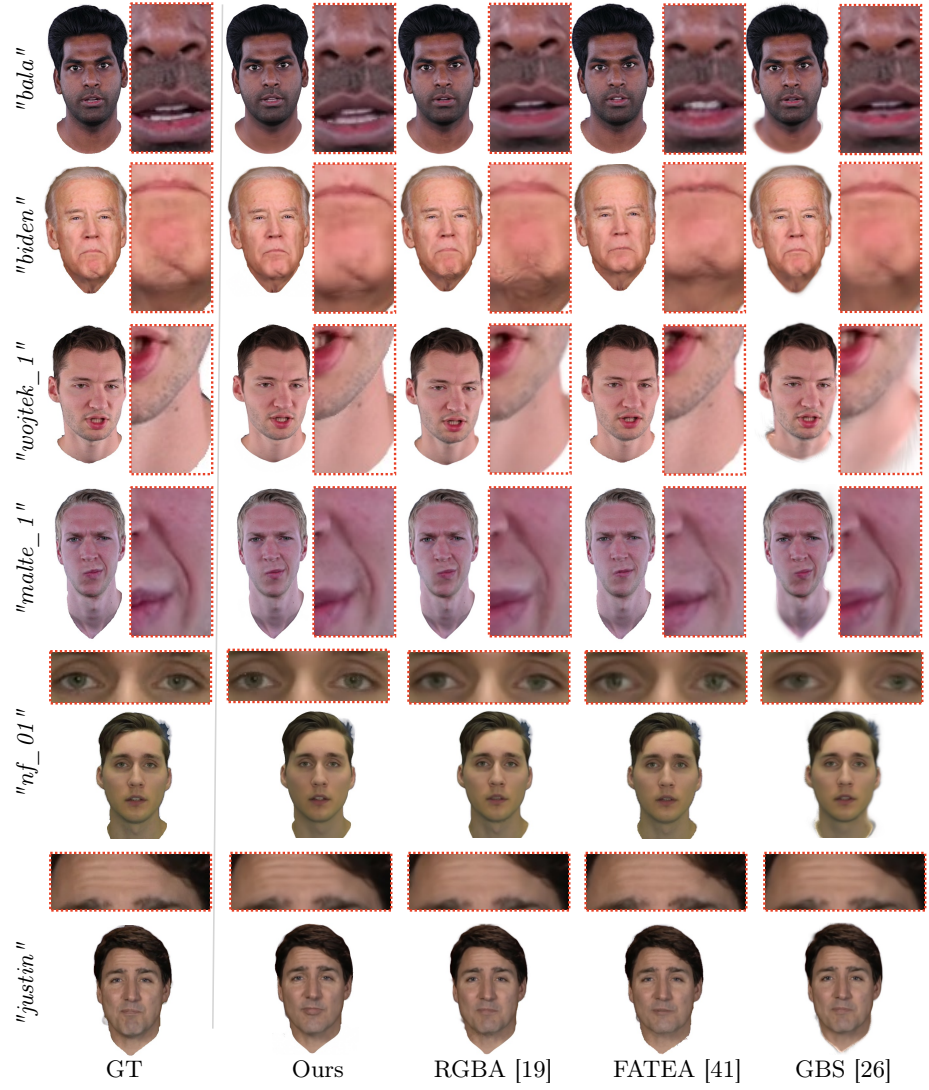
\begin{figure}
    \centering
    \begin{tikzpicture}
    \node[anchor=south west,inner sep=0] (img) at (0,0){\includegraphics[width=1.00\linewidth]{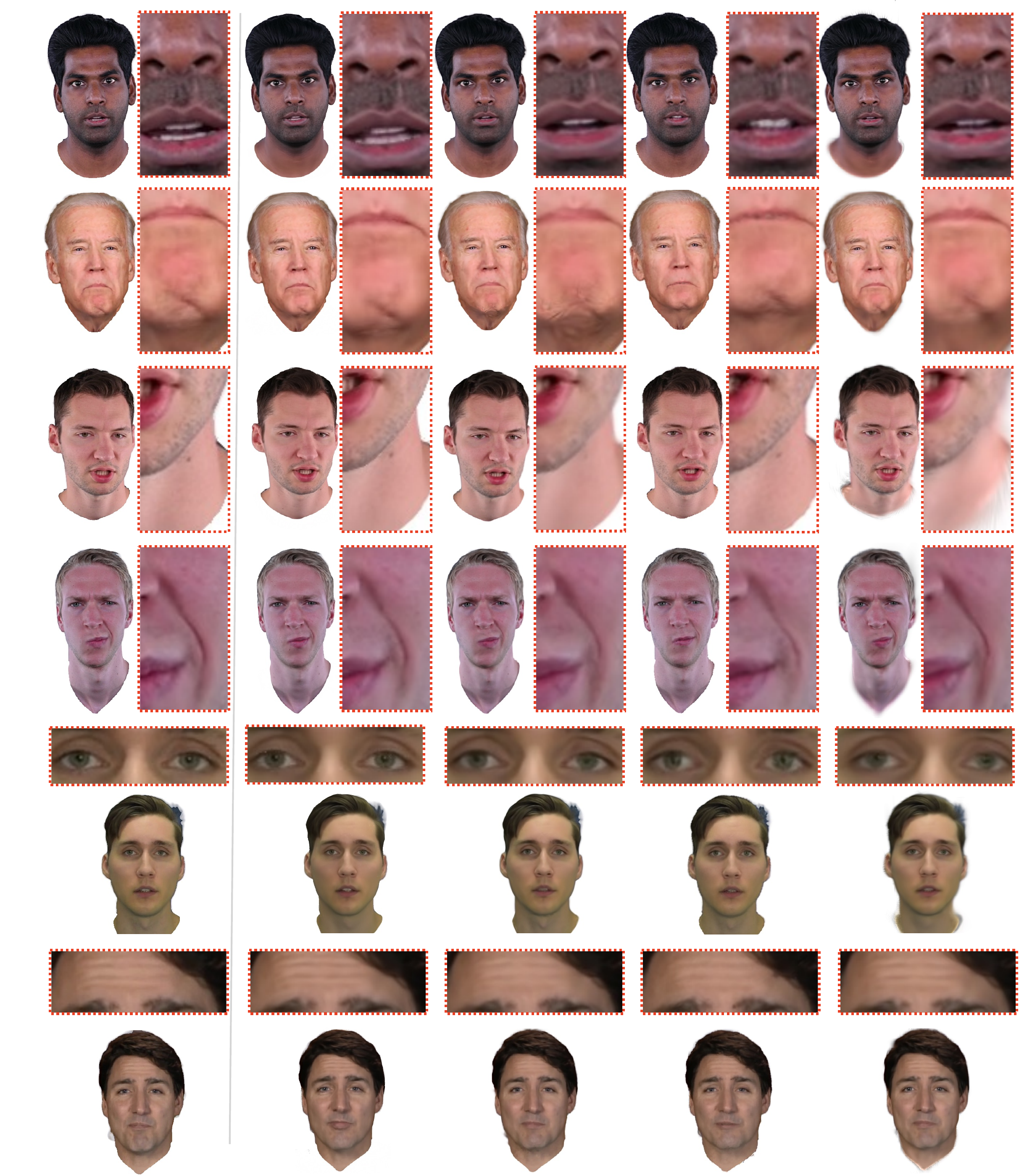}};
    \node[anchor=south] at (0.13\textwidth, -0.04\textwidth) {\strut \small GT};
    \node[anchor=south] at (0.34\textwidth, -0.04\textwidth) {\strut \small Ours };
    \node[anchor=south] at (0.53\textwidth, -0.04\textwidth) {\strut \small RGBA~\cite{rgbavatar} };
    \node[anchor=south] at (0.73\textwidth, -0.04\textwidth) {\strut \small FATEA~\cite{fate}};
    \node[anchor=south] at (0.91\textwidth, -0.04\textwidth) {\strut \small GBS~\cite{gaussianblendshape}};
    \node[anchor=south, rotate=90] at (0.05\textwidth, 1.06\textwidth) {\strut \small \textit{"bala"}};
    \node[anchor=south, rotate=90] at (0.05\textwidth, 0.89\textwidth) {\strut \small \textit{"biden"}};
    \node[anchor=south, rotate=90] at (0.05\textwidth, 0.71\textwidth) {\strut \small \textit{"wojtek\_1"}};
    \node[anchor=south, rotate=90] at (0.05\textwidth, 0.54\textwidth) {\strut \small \textit{"malte\_1"}};
    \node[anchor=south, rotate=90] at (0.05\textwidth, 0.33\textwidth) {\strut \small \textit{"nf\_01"}};
    \node[anchor=south, rotate=90] at (0.05\textwidth, 0.08\textwidth) {\strut \small \textit{"justin"}};
    \end{tikzpicture}
     \vspace{-7mm}
    \caption{\textbf{Qualitative comparisons.}   Our method produces photorealistic renderings with accurate fine details (teeth, wrinkles, skin dot, eyes) compared to four baselines. Zoom-in views highlight our detail preservation.}
   \vspace{-7mm}
    \label{fig:qualitative}
\end{figure}

\begin{figure} [t]
    \centering
    \begin{tikzpicture}
    \node[anchor=south west,inner sep=0] (img) at (0,0)
    {\includegraphics[width=0.9\linewidth]{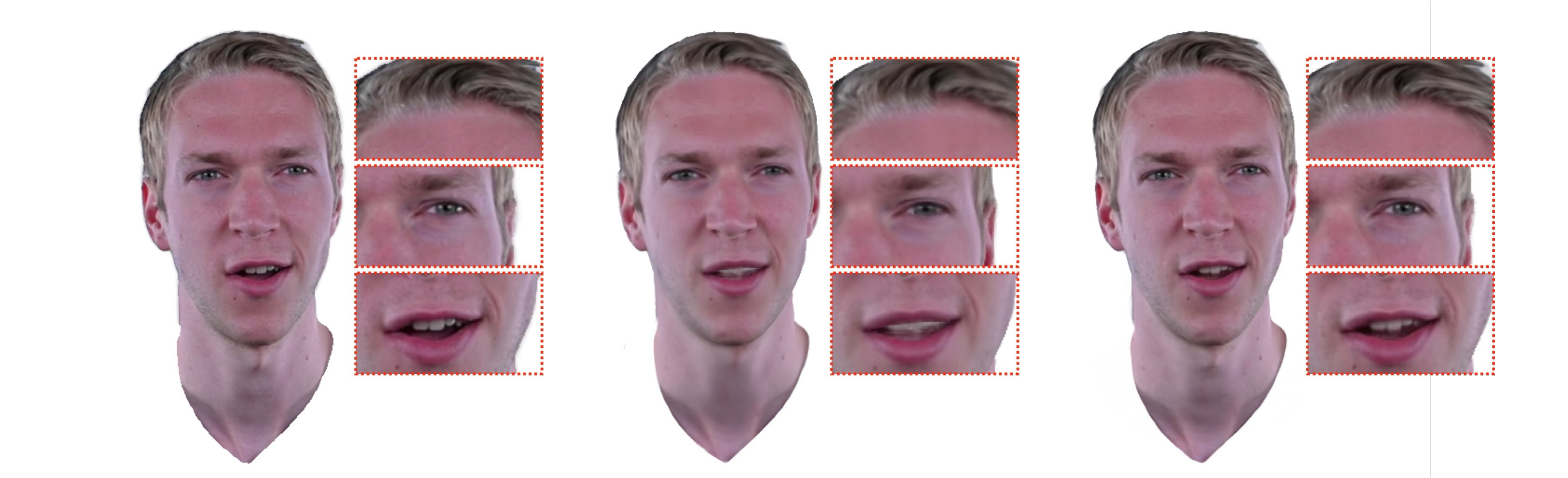}};;
    \node[anchor=south west] at (0.12\textwidth, -0.03\textwidth) {\strut \small GT};
    \node[anchor=south west] at (0.40\textwidth, -0.03\textwidth) {\strut \small Mesh};
    \node[anchor=south west] at (0.65\textwidth, -0.03\textwidth) {\strut \small  Mesh+3GS};
    \end{tikzpicture}
    \vspace{-5mm}
    \caption{\textbf{Comparison of mesh only and full model.} 
    Without Gaussians (Mesh only), the rendering lacks fine details such as teeth visibility and skin texture. Our full model (Mesh+3GS) captures these high-frequency details through Gaussian residuals.}
     \vspace{-3mm}
    \label{fig:mesh+gs}
    
\end{figure}
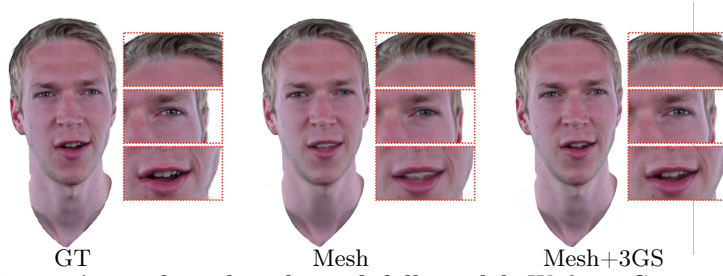

\begin{figure}[t]
    \centering
    \begin{tikzpicture}
    \node[anchor=south west,inner sep=0] (img) at (0,0){\includegraphics[width=1.0\linewidth]{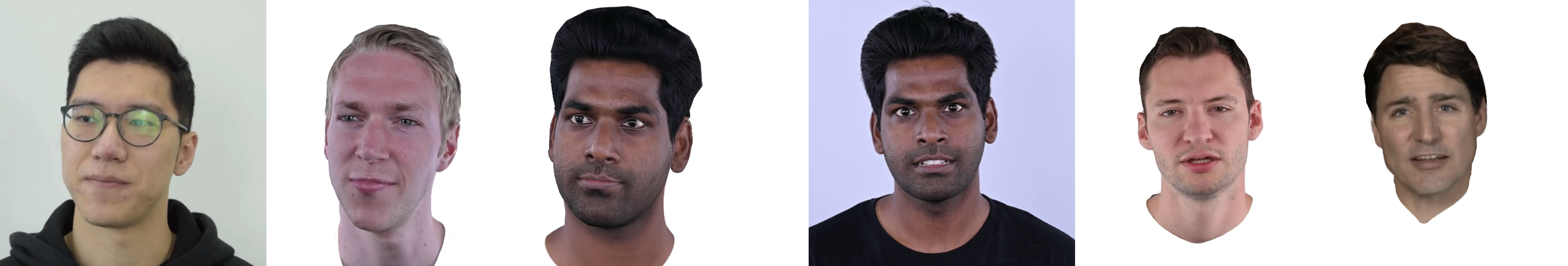}};
    \node[anchor=south] at (0.074\textwidth, -0.04\textwidth) {\strut \small Target};
    \node[anchor=south] at (0.35\textwidth, -0.04\textwidth) {\strut \small  Sources};
    \node[anchor=south] at (0.59\textwidth, -0.04\textwidth) {\strut \small Target};
    \node[anchor=south] at (0.835\textwidth, -0.04\textwidth) {\strut \small  Sources};
    \end{tikzpicture}
    \vspace{-8mm}
    \caption{\textbf{Cross-subject reenactment.} The visualization shows avatars trained on one identity (sources) are driven by facial expressions from different subjects (target).   
    Multiple source identities are driven by expressions from a single target subject.   }
    \vspace{-7mm}
\label{fig:cross_reenactment}
\end{figure}

\subsection{Qualitative Comparisons}\label{sec:qualitative}
\cref{fig:qualitative} shows the qualitative comparison with baselines.
In the \textit{bala} sequence, our method faithfully reconstructs teeth,
while baselines either blur or hallucinate the oral region. For
\textit{biden}, RGBA~\cite{rgbavatar} and FATEA~\cite{fate}
over-exaggerate facial features, while
GBS~\cite{gaussianblendshape} under-represents them due to the limited expressiveness of the predefined blendshape space. In \textit{wojtek\_1}, our method is the only one that captures facial moles, and the \textit{malte\_1} sequence shows that ours reproduces subtle wrinkles more accurately. In \textit{nf\_01}, baseline methods produce noticeably blurred renderings, whereas ours preserves vivid appearance. Finally, in \textit{justin}, our method reconstructs forehead wrinkles with temporal consistency, while RGBA~\cite{rgbavatar} partially captures forehead wrinkles but lacks overall consistency in wrinkle rendering across frames. These results suggest that our unified UV-space formulation effectively captures fine-grained details through adaptive Gaussian sampling while maintaining structural coherence via the mesh component.  Full comparisons across all subjects are provided in the supplementary material.
\cref{fig:mesh+gs} shows that the mesh can capture base geometry, but the results lack fine details (\eg teeth, skin texture), while our model recovers them via Gaussian residuals, underscoring the value of our hybrid design.

\begin{table}[pt]
    \centering
        \caption{\textbf{Ablation study of our model} on the INSTA dataset~\cite{INSTA}.}
    \label{Ablation : module}
        \vspace{-2mm}
    \scriptsize
    \setlength{\tabcolsep}{20pt}
    \renewcommand{\arraystretch}{0.7}
    \begin{tabular}{l|lll}
    \toprule
    Method                        & PSNR   & SSIM   & LPIPS   \\ 
    \midrule
    w/o 3DGS           & 33.60& 0.943 & 0.057 \\
    w/o $\mathcal{L}_{\text{normal}}$           & 34.45& 0.948 & 0.047\\ 
    w/o occlusion-aware blending  & 33.63& 0.949 & 0.045 \\
    w/o $0.2\cdot \|\mathbf{I}_{\text{mesh}} - \mathbf{I}_{\text{gt}}\|_1$  & 34.90& 0.955 & 0.031 \\
    \textbf{Ours} &  \textbf{35.31}&  \textbf{0.958}& \textbf{0.022} \\ 
    \bottomrule
    \end{tabular}
    \vspace{-2mm}
\end{table}

\noindent
\textbf{Cross-Subject Reenactment.}
\cref{fig:cross_reenactment} evaluates the generalization capability of our method through cross-subject reenactment experiments. Our method successfully transfers the facial expressions of target while preserving the identity-specific features of source. The unified UV-space representation enables smooth expression transfer without artifacts. 

\vspace{-6mm}
\subsection{Ablation Studies}\label{sec:ablation}
\vspace{-2mm}

\begin{table}[pt]
    \centering
        \caption{\textbf{Design validation of sampling strategies} by changing UV resolutions.}
    \label{Ablation : uniform sampling}
            \vspace{-2mm}
    \scriptsize
    \setlength{\tabcolsep}{10pt}
    \renewcommand{\arraystretch}{0.7}
    \begin{tabular}{c|ccc|c|ccc}
        \toprule
        Uniform  & PSNR & SSIM & LPIPS & Adaptive  & PSNR & SSIM & LPIPS\\
        \midrule
        128$^2$ & 33.49& 0.9447& 0.047 & 256$^2$ & 34.00& 0.948& 0.042\\
        256$^2$ & 33.94& 0.9482& 0.038 & 512$^2$ & 34.15& 0.949& 0.040\\
        512$^2$ & \textbf{34.11}& \textbf{0.949}& \textbf{0.034} & 1024$^2$ & 34.17& 0.949& 0.0412\\
        1024$^2$ & 33.49& 0.948& 0.0370& 2048$^2$ & \textbf{34.37}& \textbf{0.950}& \textbf{0.031}\\
        \bottomrule
    \end{tabular}
    \vspace{-3mm}
\end{table}

\begin{table}
    \centering
    \caption{\textbf{Frame-to-frame LPIPS consistency} on the INSTA dataset.}
    \vspace{-3mm}
    \scriptsize
    \setlength{\tabcolsep}{15pt}
    \begin{tabular}{lcccc}
        \toprule
         & FATEA & GBS & RGBA & \textbf{Ours} \\
        \midrule
        LPIPS Mean $\downarrow$ & 0.0325 & 0.0312 & 0.0267 & \textbf{0.0153} \\
        LPIPS Var $\downarrow$  & 0.000026 & 0.000033 & 0.000032 & \textbf{0.000010} \\
        \bottomrule
        \vspace{-9mm}
    \end{tabular}
    \label{tab:temporal_stability}
\end{table}

\noindent\textbf{Component Analysis.}
\cref{Ablation : module} analyzes the contribution of each component in our framework. At the first row, the model without 3DGS causes the most significant degradation in perceptual quality, confirming that the fixed-topology mesh alone cannot capture high-frequency appearance details. At the second row, without $\mathcal{L}_{\text{normal}}$, geometric accuracy degrades as the displacement map loses guidance from the pseudo ground-truth normals. At the third row, disabling occlusion-aware blending allows Gaussians behind the mesh surface to be rendered redundantly, introducing visual inconsistencies in the composited image. At the fourth row, removing the $0.2\cdot \|\mathbf{I}_{\text{mesh}} - \mathbf{I}_{\text{gt}}\|_1$ weakens the mesh's incentive to learn meaningful texture and geometry on its own, forcing the Gaussians to compensate for both base appearance and residual details, which reduces overall perceptual quality. We note that the model without 3DGS can occasionally surpass other models on certain subjects, as Gaussians may introduce geometric disorder; we analyze this further below.

\noindent\textbf{Uniform vs. Adaptive Sampling.}
\cref{Ablation : uniform sampling} compares uniform and adaptive sampling strategies across UV resolutions. With uniform sampling, performance peaks at $512 \times 512$, but degrades at $1024 \times 1024$, indicating overfitting from redundant Gaussians. In contrast, adaptive sampling shows improvement with increasing resolution, validating that our adaptive gaussian sampling efficiently concentrates Gaussians where most needed.

\noindent\textbf{Stability of Animation-Driven Deformation.}
To verify animation stability, we report frame-to-frame LPIPS, defined as 
$\frac{1}{N-1}\sum_{t=1}^{N-1}\text{LPIPS}(\hat{I}_t, \hat{I}_{t+1})$,
where $\hat{I}_t$ is the predicted frame at time $t$, on the INSTA dataset. As shown in \cref{tab:temporal_stability}, our method achieves the lowest frame-to-frame LPIPS mean (0.0153) and variance (0.000010), indicating that our unified UV-space representation produces temporally consistent renderings under animation-driven deformation. The low variance further confirms that our method maintains stable rendering quality across frames without introducing flickering or temporal artifacts.

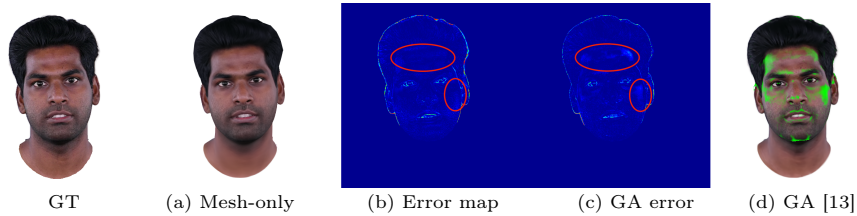
\begin{figure} 
\vspace{-6mm}
    \centering
    \begin{tikzpicture}
    \node[anchor=south west,inner sep=0] (img) at (0,0)
    {\includegraphics[width=1.0\linewidth]{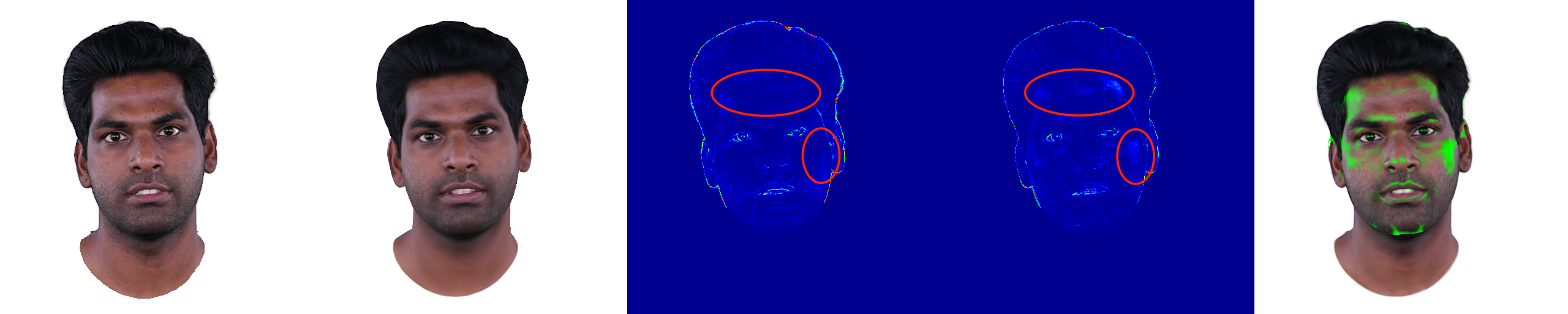}};
    \node[anchor=south] at (0.10\textwidth, -0.04\textwidth) {\strut \scriptsize GT};
    \node[anchor=south] at (0.28\textwidth, -0.04\textwidth) {\strut \scriptsize (a) Mesh-only};
    \node[anchor=south] at (0.50\textwidth, -0.04\textwidth) {\strut \scriptsize (b) Error map};
    \node[anchor=south] at (0.72\textwidth, -0.04\textwidth) {\strut \scriptsize (c) GA error};
    \node[anchor=south] at (0.9\textwidth, -0.04\textwidth) {\strut \scriptsize (d) GA~\cite{gaussianavatar}};
    \end{tikzpicture}
    \vspace{-7mm}
    \caption{\textbf{Geometric disorder in Gaussian-based baselines.} (a) Mesh-only rendering from our method. (b) Corresponding error map. (c) Error map of a GA, representative of a common failure in mesh-free Gaussian methods. (d) Rendering from GA with a green mesh embedded inside the head to expose opacity collapse. Visible green regions indicate Gaussians failing to cover the facial surface.}
    \vspace{-5mm}
    \label{fig:error_mesh_vs_gs}
\end{figure}
\noindent\textbf{Geometric Disorder.}
\label{ablation:geometric disorder}
\cref{fig:error_mesh_vs_gs} visualizes a failure common to Gaussian-based methods, exemplified by GA. To visualize this phenomenon, we embed a green mesh inside the head; when Gaussians converge to low opacity, the underlying mesh becomes visible through the face as green regions shown in (d). This occurs because Gaussians fail to place properly on the facial surface and instead scatter behind the head. In contrast, our mesh component provides a reliable geometric prior that prevents such disorder.
\vspace{-3mm}
\section{Conclusion}
\vspace{-2mm}
We have proposed URHead, a unified approach that combines mesh geometry and Gaussian appearance modeling in UV space for photorealistic 3D head avatar reconstruction. This is the first attempt to jointly optimize the mesh and Gaussian representations in the shared UV-parameterized space. This shared representation leverages mesh stability for base geometry while Gaussians capture details with adaptive sampling. Extensive experiments demonstrate competitive performance, validating the effectiveness of our unified approach. We believe that these shared representations between different renderers open new directions for combining classical and neural rendering in avatar generation to take advantage of different renderers.


\vspace{-2mm}
\section*{Acknowledgements}
\vspace{-2mm}
This work was supported by the Institute of Information \& communications Technology Planning \& Evaluation (IITP) grant funded by the Korea government(MSIT) (RS-2021-II211341, RS-2024-00398830, RS-2024-00456709).

%
%
\bibliographystyle{splncs04}
\bibliography{main}

\makeatletter
\providecommand{\thefootnote}{}
\makeatother

\title{Supplementary Material}
\author{}
\institute{}

\maketitle

\setcounter{section}{0} 
\setcounter{figure}{0}
\renewcommand{\thefigure}{A\arabic{figure}}

\setcounter{table}{0}
\renewcommand{\thetable}{A\arabic{table}}

\setcounter{equation}{0}
\renewcommand{\theequation}{A\arabic{equation}}

\section{UV Feature Map Visualization}
    \renewcommand\twocolumn[1][]{#1}%
    \begin{center}
    \begin{minipage}[t]{\linewidth}
          \centering
          \includegraphics[width=1.0\linewidth]{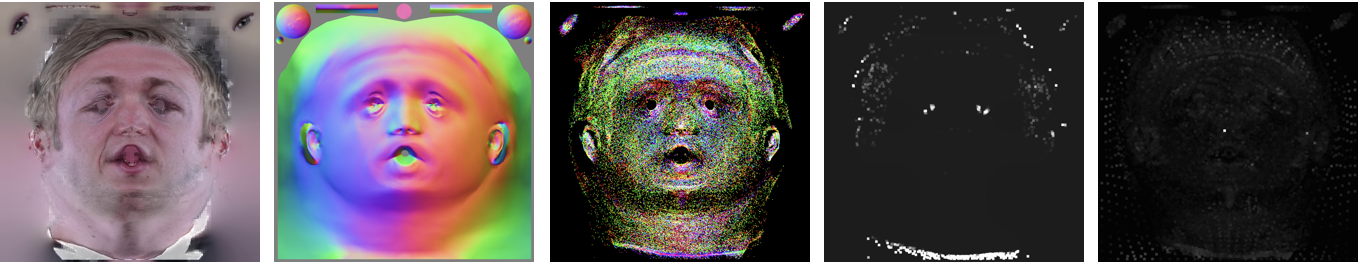}
      \end{minipage}
    \begin{minipage}[t]{0.19\linewidth}
          \centering
           \small Color $\textbf{C}$
      \end{minipage}
    \begin{minipage}[t]{0.19\linewidth}
          \centering
           \small  Normal $\textbf{N}$
      \end{minipage}
    \begin{minipage}[t]{0.19\linewidth}
          \centering
           \small Scale $\textbf{S}$
      \end{minipage}
    \begin{minipage}[t]{0.19\linewidth}
          \centering
           \small Opacity $\textbf{O}$
      \end{minipage}
    \begin{minipage}[t]{0.19\linewidth}
          \centering
           \small Displacement $\textbf{D}$
      \end{minipage}
    \captionof{figure}{\textbf{Visualization of learned UV feature maps.} Our unified UV representation encodes multiple attributes of 3DGS: color map $\mathbf{C}$ captures appearance, normal map $\mathbf{N}$ encodes orientation, scale map $\mathbf{S}$ controls Gaussian size, opacity map $\mathbf{O}$ determines visibility, and displacement map $\mathbf{D}$ adjusts positions.}
    \label{fig:uv_features}
    \end{center}
Figure~\ref{fig:uv_features} visualizes the learned UV feature maps that encode our unified representation. Each map stores specific Gaussian attributes: the color map $\mathbf{C}$ captures RGB appearance with smooth variations across facial regions; the normal map $\mathbf{N}$ encodes surface orientation; the scale map $\mathbf{S}$ reveals adaptive Gaussian scales with higher density in detail regions like eyes and mouth; the opacity map $\mathbf{O}$ controls visibility with higher values on the face surface and lower values at boundaries; and the displacement map $\mathbf{D}$ stores geometric corrections to the base mesh. This UV-based parameterization enables efficient storage and sampling of all Gaussian attributes in a spatially coherent manner.

\section{Joint optimization of meshes and Gaussians.}
\label{reb:joint optimization}
We 
clarify that mesh \textit{geometry} is jointly optimized, not provided by an offline FLAME predictor. The full pipeline is:
\vspace{-2mm}
\begin{equation*}
\{\boldsymbol{\psi},\boldsymbol{\theta},\mathbf{t}\}\!\xrightarrow{\text{LBS}}\!\mathbf{v}_i\!\xrightarrow{\mathbf{D}}\!\hat{\mathbf{v}}_i=\mathbf{v}_i+\mathbf{D}(u_i,v_i)\!\cdot\!\mathbf{N}(u_i,v_i),
\vspace{-2mm}
\end{equation*}
where (i)~the FLAME parameters $\{\boldsymbol{\psi},\boldsymbol{\theta},\mathbf{t}\}$ are \textit{learnable} and (ii)~$\mathbf{D}$ is a \textit{learnable} UV map that physically moves vertices along normals every iteration (Eq. \ref{rq:position}). Both renderers share the same displaced $\hat{\mathcal{V}}$:
\vspace{-2mm}
\begin{align*}
\mathbf{I}_{\text{mesh}},d_{\text{mesh}} &= \mathcal{R}_{\text{mesh}}(\mathbf{C},\mathbf{N},\mathbf{D},\hat{\mathcal{V}},\mathcal{F}),\\
\mathbf{I}_{\text{gs}},T &= \mathcal{R}_{\text{gs}}(\mathbf{C},\mathbf{N},\mathbf{O},\mathbf{S},d_{\text{mesh}},\mathcal{G}(\hat{\mathcal{V}})).
\vspace{-6mm}
\end{align*}
\noindent Hence for any geometric parameter $\Theta\!\in\!\{\boldsymbol{\psi},\boldsymbol{\theta},\mathbf{t},\mathbf{D}\}$,
\vspace{-2mm}
\begin{equation*}
\frac{\partial\mathcal{L}_{\text{total}}}{\partial\Theta}
=\underbrace{\frac{\partial\mathcal{L}_{\text{photo}}}{\partial\mathbf{I}_{\text{mesh}}}\frac{\partial\mathbf{I}_{\text{mesh}}}{\partial\hat{\mathcal{V}}}\frac{\partial\hat{\mathcal{V}}}{\partial\Theta}}_{\text{mesh flow}}
+\underbrace{\frac{\partial\mathcal{L}_{\text{photo}}}{\partial\mathbf{I}_{\text{gs}}}\frac{\partial\mathbf{I}_{\text{gs}}}{\partial\hat{\mathcal{V}}}\frac{\partial\hat{\mathcal{V}}}{\partial\Theta}}_{\text{Gaussian flow}},
\end{equation*}

\noindent so mesh geometry is refined by gradients from \textit{both} renderers. 
The hair/accessories limitation is unrelated to the geometry pipeline; it arises because they are not covered by the FLAME mesh and thus have no UV coverage for $\mathbf{D}$.

\section{VHAP Modification}
\begin{figure}
    \centering
    \includegraphics[width=0.9\linewidth]{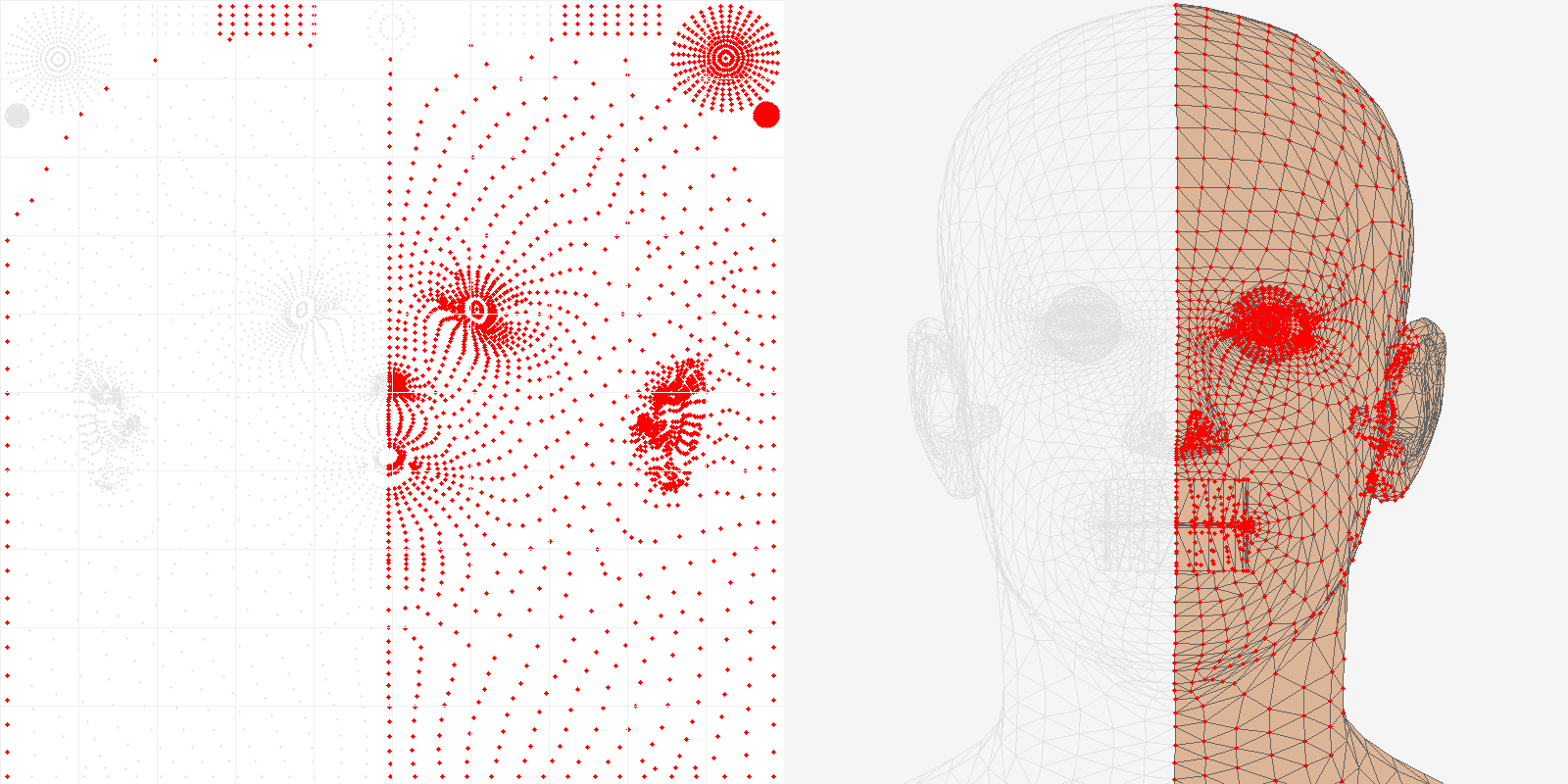}
    \caption{\textbf{UV-to-3D correspondence visualization.} Left: UV space representation of the left facial half showing vertex distribution. Right: Corresponding 3D mesh region demonstrating the spatial mapping between UV coordinates and facial geometry.}
    \label{fig:placeholder}
\end{figure}
\noindent\textbf{Revised Teeth UV Mapping.} We refine the teeth UV parameterization of VHAP~\cite{vhap} by adjusting the vertex ordering and introducing spatial offsets to separate upper and lower teeth. This modification eliminates rendering artifacts and texture discontinuities present in the original procedural grid generation. Furthermore, we introduce a simple mouth plate geometry that serves as an occlusion socket, effectively preventing visibility of the posterior cranial region through the oral cavity during rendering. Figure~\ref{fig:placeholder} visualizes the spatial correspondence between UV space and 3D geometry for a facial region.

\noindent\textbf{Seamless UV Parameterization}
To enable direct sampling of vertex parameters from UV textures, we eliminate UV seams by mapping each 3D vertex to a unique UV coordinate based on its first occurrence in the mesh topology. This creates a bijective vertex-UV mapping that allows seamless sampling of displacement, normals, and other per-vertex attributes from our unified UV representation.

\section{FPS and number of Gaussian points.}
URHead reaches 62 FPS at $512{\times}512$ on a single RTX~3090. URHead uses the second fewest Gaussians while attaining the best PSNR and LPIPS on \textit{bala} (INSTA).
The shared UV representation reduces Gaussian redundancy by allowing the mesh branch to model low-frequency structure while Gaussians focus on residual details.
\definecolor{best}{RGB}{255, 142, 142}
\definecolor{second}{RGB}{255, 204, 153}
\begin{table}
    \caption{\textbf{FPS, Gaussian count, and rendering quality comparison on \textit{bala} (INSTA).}}
    \centering
    \setlength{\tabcolsep}{7pt}
    \begin{tabular}{lccccccc}
    \toprule
    & SA & GA & FA & GBS & FATEA & RGBA & \textbf{Ours} \\
    \midrule
    \#GS                & 361K   & 58K    & \cellcolor{best}13K  & 73K    & 35K    & 60K    & \cellcolor{second}{24K} \\
    PSNR$\uparrow$      & 32.66  & 33.98  & 31.50       & 34.70  & 31.62  & \cellcolor{second}{35.15} & \cellcolor{best}{35.31} \\
    LPIPS$\downarrow$   & 0.0325 & 0.0445 & 0.0407      & 0.0563 & 0.0365 & \cellcolor{second}{0.0226} & \cellcolor{best}{0.0149} \\
    \bottomrule
    \end{tabular}
\end{table}
\vspace{-10mm}
\section{Sensitivity analysis on loss weighting.}
We show the additional $(\ell_1, \ell_1^{\text{mesh}})$ ratios on \textit{bala} (INSTA).
Both (b) and (c) improve over (a) across all metrics. This shows that the mesh loss is the key factor, and our method is robust to the ratio. Still, over-weighting the mesh term  pushes the FLAME topology toward a low-frequency approximation and limits high-frequency detail, which raises LPIPS in (c), (d), (e). 
Thus, we adopt (b) for the best perceptual quality.
\begin{table}
    \caption{\textbf{Sensitivity analysis on loss weighting ratios on \textit{bala} (INSTA).}}
    \centering
    \setlength{\tabcolsep}{8pt}
    \begin{tabular}{lccccc}
    \toprule
    ratio & (a)1.0/0 & \textbf{(b)0.8/0.2} &  (c)0.5/0.5 &  (d)0.2/0.8 &  (e)0.0/1.0 \\
    \midrule
    PSNR$\uparrow$  & 34.90 & \cellcolor{second}35.31 & \cellcolor{best}{35.38} & 	34.66 & 34.56\\
    SSIM$\uparrow$  & 0.955 & \cellcolor{second}0.958 & \cellcolor{best}{0.959} & 0.954 & 0.954\\
    LPIPS$\downarrow$ & \cellcolor{second}0.031 & \cellcolor{best}{0.022} & 0.037 &  0.043 & 0.044\\
    \bottomrule
    \end{tabular}
\end{table}

\section{Generalization Abilities.} We transfer our pre-trained head Gaussians to SMPL-X by rigidly aligning the FLAME and SMPL-X head templates. It produces a full body avatar without retraining. The head detail is preserved as SMPL-X, confirming the generalizability of ours.
\begin{figure} [h!]
\vspace{-3mm}
    \centering
    \includegraphics[width=1.0\linewidth]{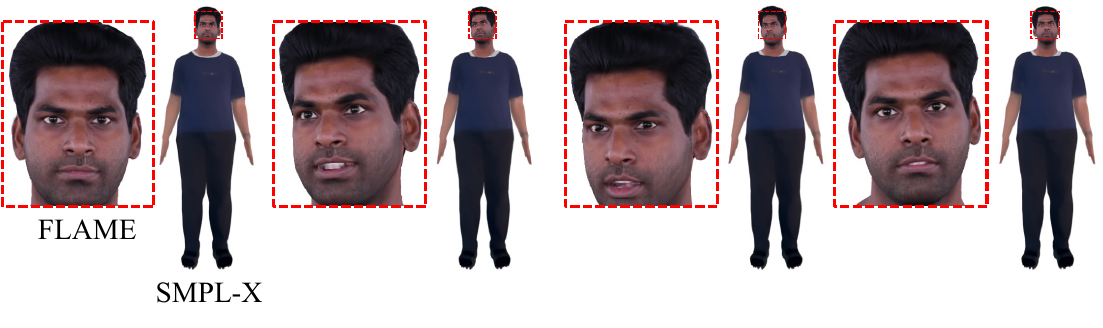}
    \label{fig:SMPL-X}
    \caption{\textbf{Generalization to full-body avatar.}}
\end{figure}

\section{Qualitative Comparison for Ours vs. Mesh only}
\begin{figure}
    \centering
    \begin{tikzpicture}
    \node[anchor=south west,inner sep=0] (img) at (0,0){\includegraphics[width=0.5\linewidth]{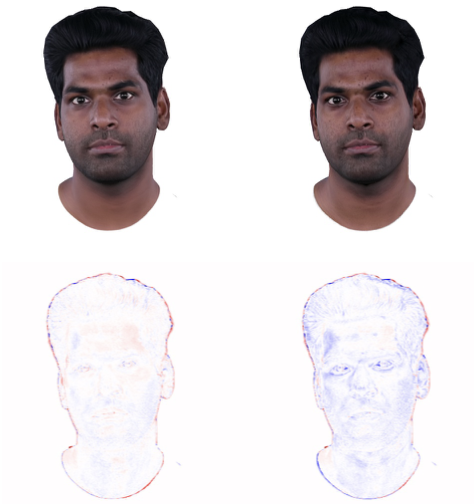}};
    \node[anchor=south] at (0.12\textwidth, 0.235\textwidth) {\strut \small Ours};
    \node[anchor=south] at (0.38\textwidth, 0.235\textwidth) {\strut \small  Mesh only};
    \end{tikzpicture}
    \caption{\textbf{Ours vs. Mesh only.} Rendering comparison (top) and error maps (bottom) demonstrate that 3D Gaussians significantly improve rendering quality over mesh-only baseline.}
    \label{fig:additional_Mesh+3DGS}
\end{figure}

Figure~\ref{fig:additional_Mesh+3DGS} compares our unified representation against mesh-only rendering to demonstrate the contribution of 3D Gaussians. While the mesh provides a strong geometric prior, rendering with the mesh alone results in visible artifacts and limited texture detail due to finite mesh resolution. By augmenting the mesh with 3D Gaussians, our method captures high-frequency appearance details and achieves photorealistic rendering quality. The error maps in the bottom row quantitatively visualize the difference from ground truth, clearly showing that our approach significantly reduces rendering error compared to mesh-only rendering, particularly in regions requiring fine-scale detail representation.

\section{Limitations and Future Work}
\label{sec:limitation}

\begin{figure}
    \centering
    \includegraphics[width=1.0\linewidth]{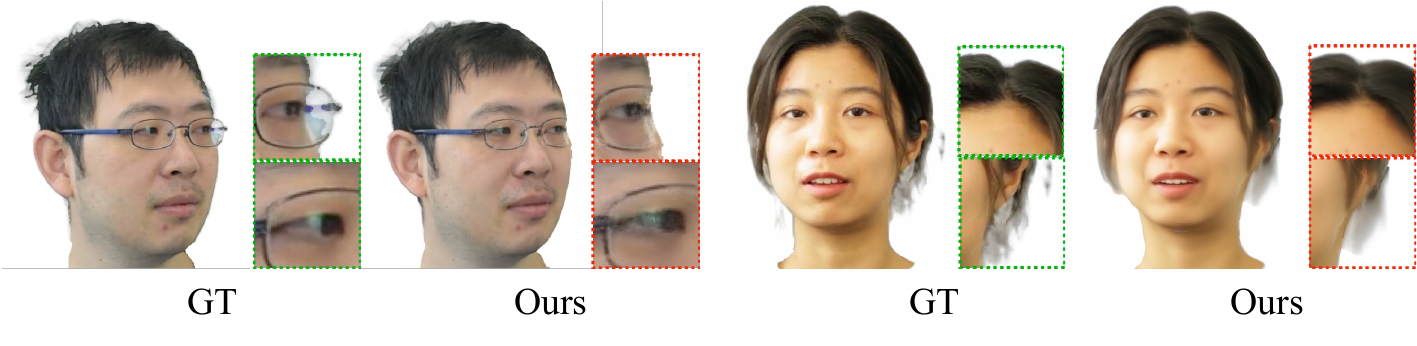}
    \caption{\textbf{Challenging cases.} Left: Thin accessories such as eyeglasses remain difficult to reconstruct precisely, showing geometric inaccuracies (green) and rendering artifacts (red). Right: Long hair extending beyond the FLAME topology cannot be explicitly modeled, resulting in incomplete reconstruction in the hair region.}
    \label{fig:limitation}
\end{figure}

As our method builds upon the FLAME parametric model, it naturally shares the scope of FLAME's topology, which primarily covers the head and upper neck regions. Regions beyond this topology, such as long hair or the torso, are not explicitly modeled. Additionally, reconstruction quality depends on the accuracy of FLAME tracking during preprocessing, which is a common requirement among FLAME-based approaches~\cite{gaussianavatar, flashavatar, rgbavatar}.

Figure~\ref{fig:limitation} illustrates two representative failure cases. First, thin accessories such as eyeglass frames are difficult to reconstruct precisely, as they frames lie outside the scope of the FLAME topology. Second, as noted in the main paper, the \textit{yufeng} subject with long hair extending beyond the FLAME topology results in incomplete reconstruction in the hair region, since our method does not explicitly model geometry outside the parametric head mesh.

\section{Additional Qualitative Results}

\begin{figure}
    \centering
    \includegraphics[width=1.0\linewidth]{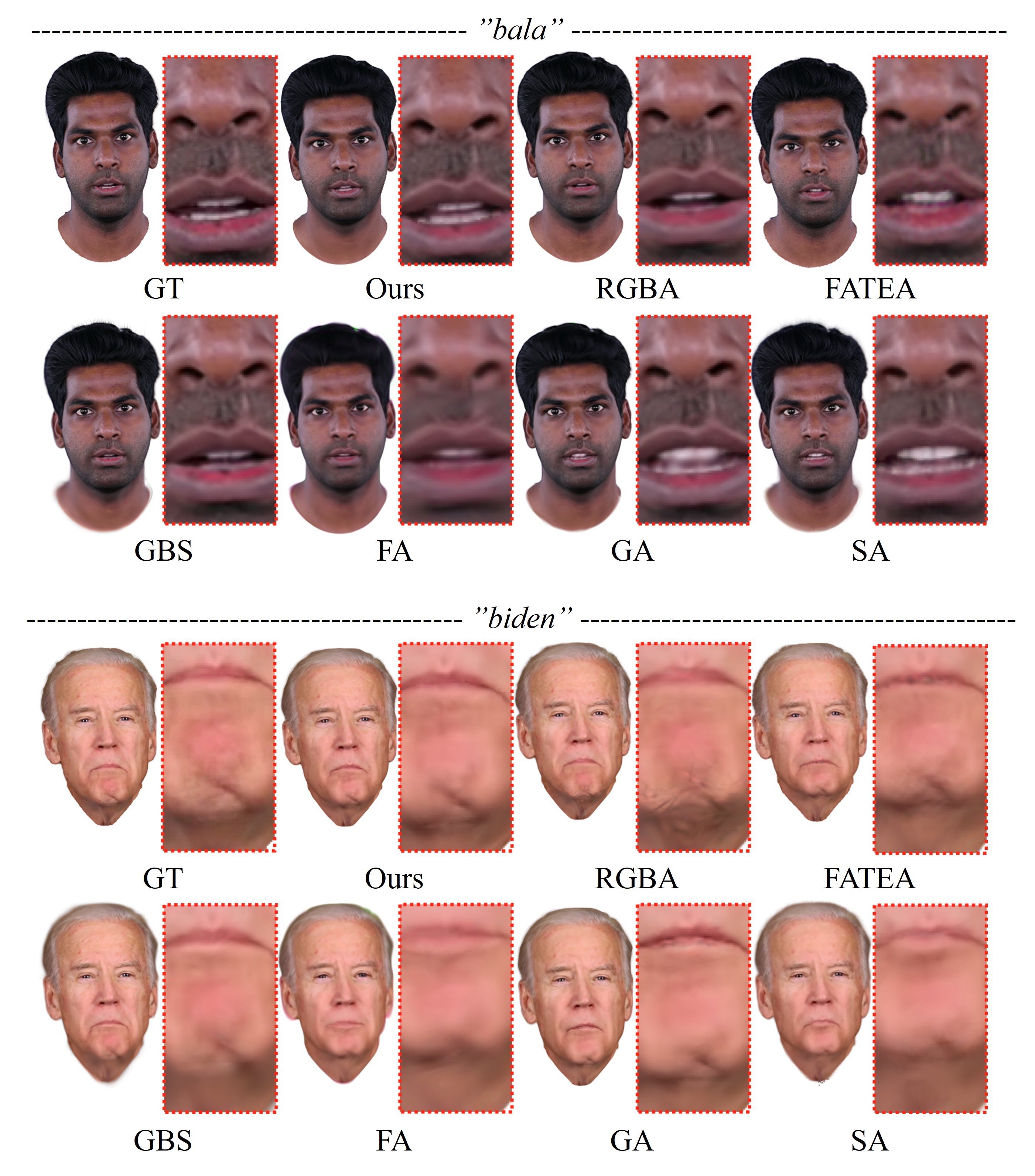}
    \caption{
        Qualitative comparisons on the \textit{bala} and \textit{biden} sequences. Our method accurately reconstructs tooth geometry and subject-specific wrinkle patterns, while baseline methods exhibit oversmoothing or geometric inaccuracies.
    }
    \label{fig:supp_bala_biden}
\end{figure}

\begin{figure}
    \centering
    \includegraphics[width=1.0\linewidth]{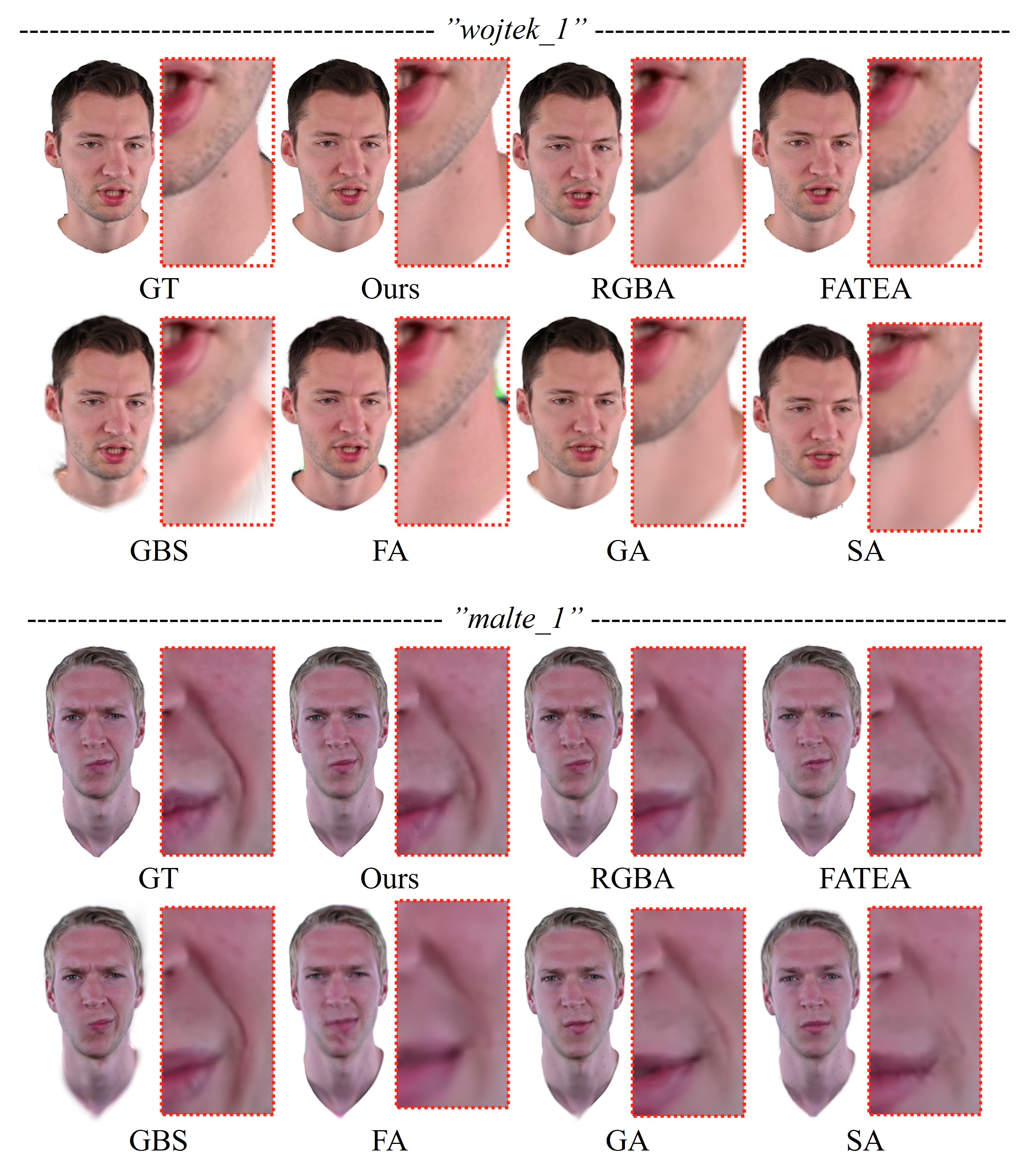}
    \caption{
        Qualitative comparisons on the \textit{wojtek\_1} and \textit{malte\_1} sequences. Our method preserves subject-specific skin features such as moles and fine wrinkles that are missed or blurred by all competing methods.
    }
    \label{fig:supp_wojtek_malte}
\end{figure}

\begin{figure}
    \centering
    \includegraphics[width=1.0\linewidth]{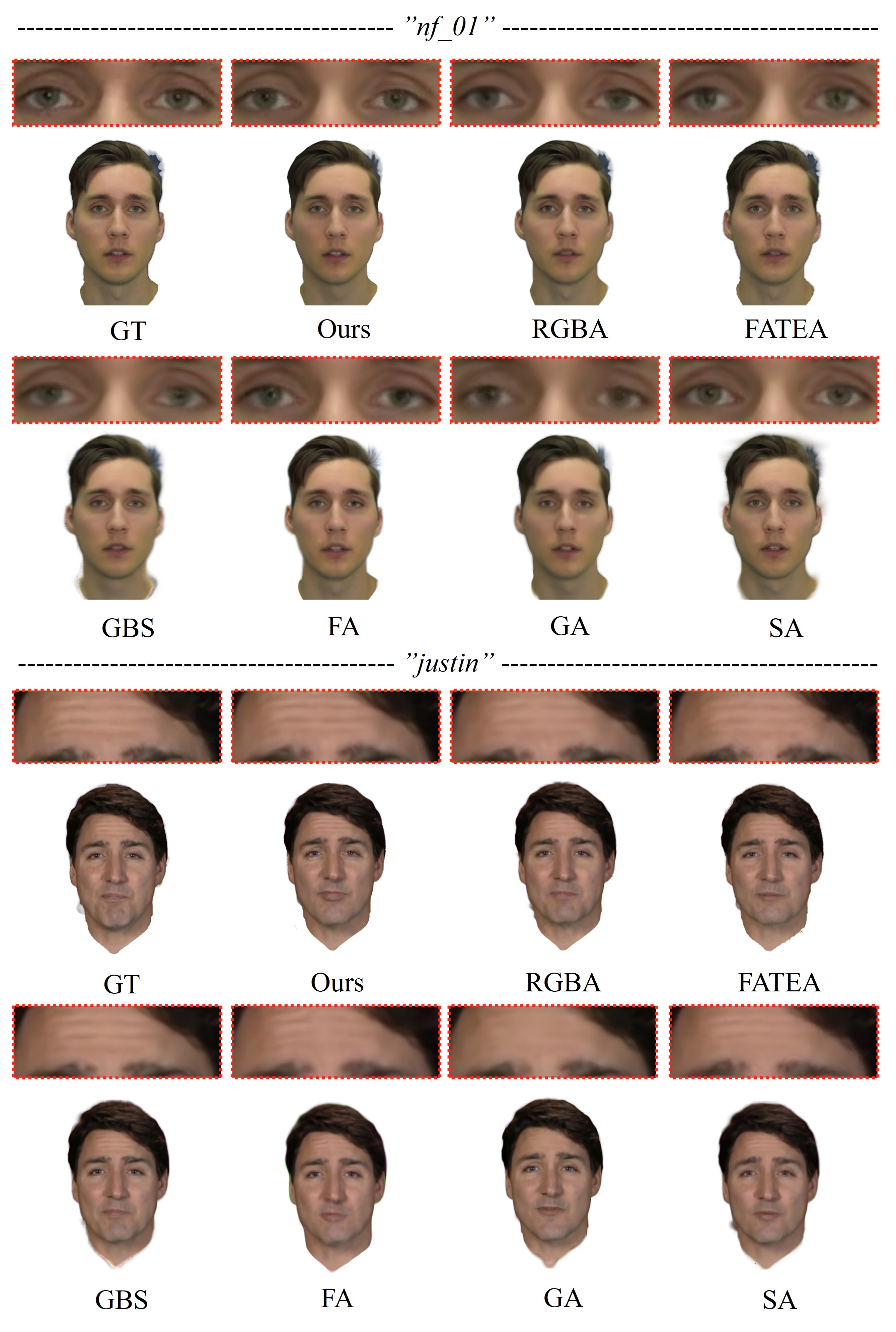}
    \caption{
        Qualitative comparisons on the \textit{nf\_01} and \textit{justin} sequences. Our method preserves fine eye structures and temporally consistent forehead wrinkles, whereas baseline methods exhibit heavy oversmoothing in both cases.
    }
    \label{fig:supp_nf_justin}
\end{figure}

We provide additional qualitative comparisons across six subjects in Figs.~\ref{fig:supp_bala_biden}--\ref{fig:supp_nf_justin}, extending the main paper by including SplattingAvatar~\cite{splattingavatar} (SA), GaussianAvatars~\cite{gaussianavatar} (GA), and FlashAvatar~\cite{flashavatar} (FA), which were omitted due to space constraints. Our method consistently produces photorealistic renderings with accurate reproduction of fine-grained details across all subjects. In \textit{bala}, our method faithfully reconstructs individual tooth geometry in the wide-open mouth region, where all baselines produce blurred or geometrically inaccurate oral cavities. In \textit{biden} and \textit{malte\_1}, our method accurately reproduces subject-specific wrinkle patterns and lip texture, while competing methods exhibit oversmoothing. In \textit{wojtek\_1}, ours is the only method that preserves the subject's skin mole and fine skin texture in the cheek region. In \textit{nf\_01}, our method retains fine eye structures including catchlights and eyelash geometry, whereas all baselines produce heavily blurred eye regions. Finally, in \textit{justin}, our method reproduces forehead wrinkles with temporal consistency across frames, while RGBA~\cite{rgbavatar} only partially captures coarse wrinkle structures and the remaining baselines yield a uniformly smooth appearance. These results corroborate our quantitative findings and further highlight the effectiveness of our unified UV-space representation in capturing subject-specific details while maintaining structural coherence.

\clearpage
\section{Additional Cross-Identity Reenactment Results}
\begin{figure}
\centering
\vspace{-7mm}
{\includegraphics[width=1.0\linewidth]{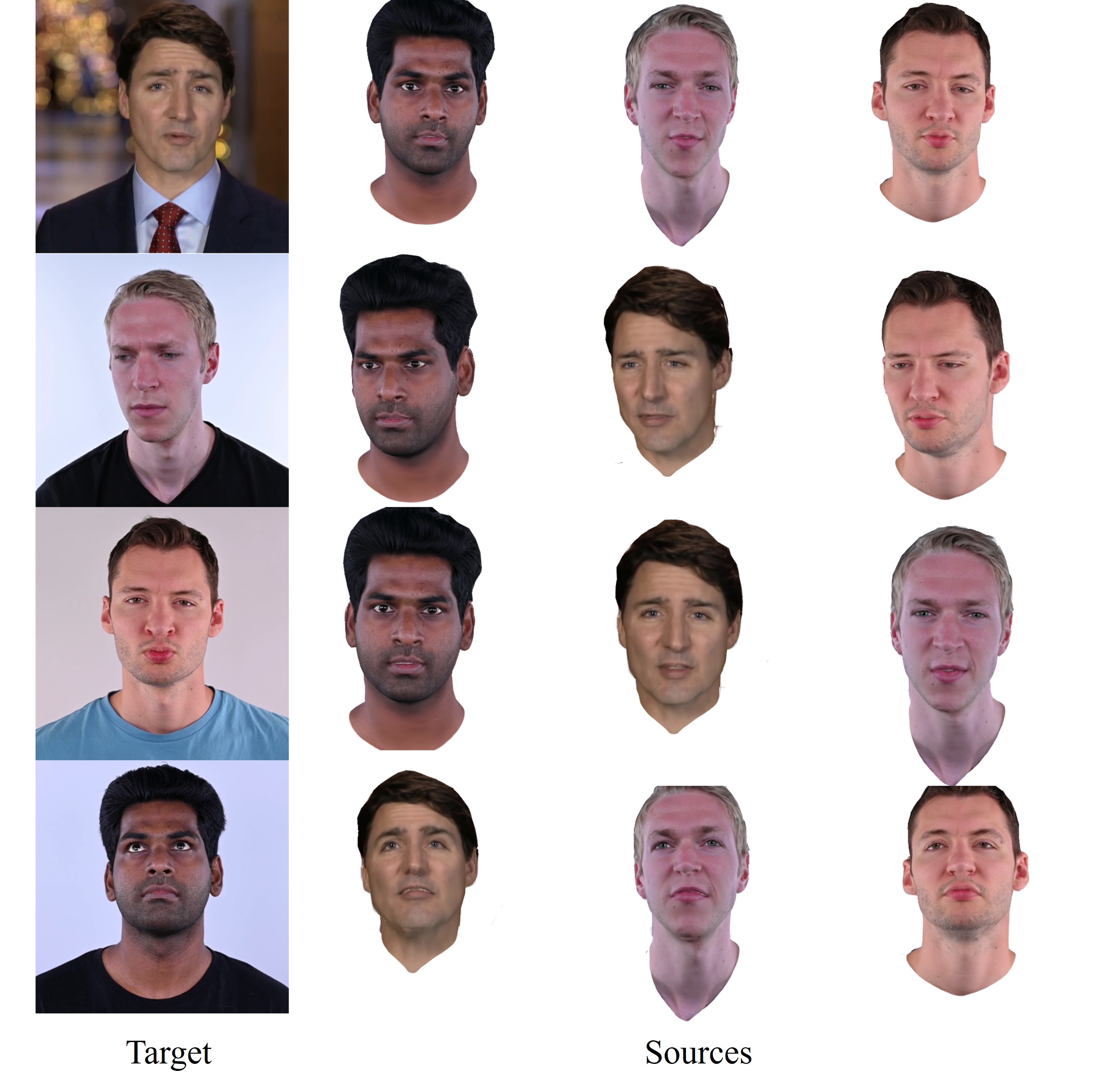}}
\caption{\textbf{Additional cross-identity reenactment.} Our method successfully drives multiple source identities (subsequent columns) using expressions from a target identity (first column), while preserving the identity-specific appearance of each source.}
\label{fig:additional_cross_reenactment}
\end{figure}

Figure~\ref{fig:additional_cross_reenactment} shows results where multiple source identities (subsequent columns) are driven by expressions from a target identity (first column). Our method successfully preserves distinctive appearance characteristics of each source identity while accurately reproducing the target expressions across different subjects. The results demonstrate consistent rendering quality with faithful reproduction of identity-specific attributes including facial structure, skin tone, and hair appearance, while maintaining precise expression transfer. These validate the robustness of our UV-based parametric representation for cross-identity animation and highlight the effective disentanglement between identity and expression in our framework.

\end{document}